\documentstyle[12pt,aasms4,psfig]{article}

\def\ltsima{$\; \buildrel < \over \sim \;$}
\def\lsim{\lower.5ex\hbox{\ltsima}}
\def\gtsima{$\; \buildrel > \over \sim \;$}
\def\gsim{\lower.5ex\hbox{\gtsima}}

\begin{document}
\title
{GRAVITATIONAL LENS STATISTICS FOR GENERALIZED NFW PROFILES: PARAMETER DEGENERACY AND IMPLICATIONS FOR SELF-INTERACTING COLD DARK MATTER}

\author{J. S. B. Wyithe\altaffilmark{1,2}, E. L. Turner\altaffilmark{1} \& D. N. Spergel\altaffilmark{1,3}}

\altaffiltext{1}{Princeton University Observatory, Peyton Hall, Princeton, NJ 08544, USA}

\altaffiltext{2}{School of Physics, The University of Melbourne, Parkville, Vic, 3052, Australia}

\altaffiltext{3}{The Institute for Advanced Study, Princeton, NJ 08540}

\begin{abstract}

Strong lensing is a powerful probe of the distribution of matter in the cores of clusters of galaxies. Recent studies suggest that the cold dark matter model predicts cores that are denser than those observed in galaxies, groups and clusters. One possible resolution of the discrepancy is that the dark matter has strong interactions (SIDM), which leads to lower central densities. A generalized form of the Navarro, Frenk and White profile (Zhao profile) may be used to describe these halos. In this paper we examine gravitational lensing statistics for this class of model. 

The optical depth to multiple imaging is a very sensitive function of the profile parameters in the range of interest for SIDM halos around clusters of galaxies. Less concentrated profiles, which result from larger self-interaction cross-sections, can produce many fewer lensed pairs. Furthermore, profiles that result in a small optical depth exhibit reduced typical splittings, but produce multiple images that are more highly magnified. However the resulting increased magnification bias does not alter our conclusions. 

We find that lensing statistics based on profile parameters obtained from fits out to the virial radius are dependent on the minimization scheme adopted, and may be seriously in error. However, profile fits weighted towards the core region have parameter degeneracies that are approximately equivalent to those for strong lensing cross-sections. 

Lensing statistics provide a powerful test for SIDM. More realistic and observationally oriented calculations remain to be done, however larger self-interaction cross-sections may well be ruled out by the very existence of strong lenses on galaxy cluster scales. The inclusion of centrally dominant cluster galaxies should boost the cross-section to multiple imaging. However our preliminary calculations suggest that the additional multiple imaging rate is small with respect to the differences in multiple imaging rate for different halo profiles. In future statistical studies, it will be important to properly account for the scatter among halo profiles since the optical depth to multiple imaging is dominated by the most concentrated members of a cluster population.

\end{abstract}


\section{INTRODUCTION}

The cold dark matter plus cosmological constant model ($\Lambda$CDM) appears to be remarkably successful in describing cosmology on scales much larger than a Mpc (e.g. Bahcall, Ostriker, Perlmutter \& Steinhardt 1999). However recent observations on smaller scales find $\Lambda$CDM wanting: $\Lambda$CDM predicts dwarf galaxy halos (e.g. Moore et al. 1999) that are denser and more concentrated than observed (Dalcanton \& Bernstein 2000; Firmani et al. 2000b). The persistence of bars in galaxies like our own, suggests that the dark matter is not centrally concentrated (Sellwood 2000). In addition, the intensively studied CL0024+1654 is nearly spherical (Colley, Tyson \& Turner 1996; Tyson, Kochanski \& Dell'antonio 1998), at odds with $\Lambda$CDM predictions of triaxial systems. Spergel \& Steinhardt (2000) have recently suggested that these and other discrepancies (see for example Dav$\acute{\mathrm{e}}$, Spergel, Steinhardt \& Wandelt 2000; Wandelt et al. 2000 and references therein) could be resolved if cold dark matter were weakly self interacting with a large scattering cross-section (SIDM). Several authors (Burkert 2000; Yoshida, Springel, White \& Tormen 2000; Dav$\acute{\mathrm{e}}$ et al. 2000) have since undertaken N-body simulations of halos with self interactions in the regime suggested by Spergel \& Steinhardt (2000). These works find that SIDM halos produce flatter, smoother and more spherical cores than their $\Lambda$CDM counterparts.

 We are interested in the statistics of multiple imaging due to a generalized NFW profile (Zhao 1996) which may be used to describe both CDM and SIDM halos. We concentrate our efforts on lensing by clusters of galaxies, since single galaxies contain a significant mass in baryons, distributed to produce a flat rotation curve. This makes SIDM in galaxies difficult to probe using gravitational lensing. Conversely, clusters of galaxies have a much larger mass to light ratio, and baryons which are less centrally concentrated (particularly in the absence of a central dominant galaxy). We assume spherically symmetric profiles and concentrate on the statistics of the optical depth to multiple imaging. The total magnification, and the maximum image splitting are also considered. 

In Secs.~\ref{darkmatter} and \ref{lensing} we discuss the properties of dark matter halos and introduce the formalism for computing the lensing statistics of the Zhao profile. In Secs.~\ref{results}, \ref{degeneracy} and \ref{implications} we discuss the statistics obtained, contrast the parameter degeneracies present in the lens statistics and in profile fitting, and present some implications for SIDM cluster halos. In Sec.~\ref{halodist} we discuss the important implications for lens statistics of the distribution of halo profile-parameters. We also discuss two potentially important caveats which turn out not to effect our conclusions: we discuss magnification bias (Sec.~\ref{amp_dist_bias}), and give a preliminary account of the effect of massive central galaxies on the multiple imaging rate (Sec.~\ref{galaxies}). Throughout the paper we have assumed a standard $\Omega_o=1$, $\Lambda=0$ and $h=\frac{H_o}{100km\,sec_{-1}}=0.7$ filled beam cosmology.

\section{DARK MATTER PROFILES}
\label{darkmatter}

Navarro, Frenk \& White (1997) claimed the NFW profile (with an inner density $\rho(r)$ varying with radius $r$ as $\rho(r)\propto r^{-1}$) to be universal. More recently Moore et al. (1999) and Ghingna et al. (2000) have claimed a different universal profile, having $\rho(r)\propto r^{-\frac{3}{2}}$ in the innermost regions. However Jing \& Suto (2000) find a systematic correlation of inner profile slope with mass. Moreover, they find some scatter in the slope at a fixed mass. On the other hand, from their convergence study, Klypin, Kravtsov, Bullock \& Primack (2000) find that while there is some scatter, in general halos are equally well fit in the region where convergence is achieved by either an NFW or a Moore profile, independent of the halo mass. They interpret the correlation found by Jing \& Suto (2000) as a trend in halo concentration rather than in asymptotic slope. Dav$\acute{\mathrm{e}}$, Spergel, Steinhardt \& Wandelt (2000) present a comprehensive suite of simulations of both CDM and SIDM dwarf galaxy halos. They find typical profiles for CDM halos in the inner most regions of $\rho(r)\sim r^{-\frac{3}{2}}$, with some scatter. However they found SIDM halos to be less cuspy, having central profiles ranging from $\rho(r)\sim r^{-1}$ to $\rho(r)\sim r^{-\frac{1}{2}}$. The uncertainty as to the optimal halo profile as well as the variation introduced by SIDM motivate an exploration of lensing by profiles that differ from the analytically treatable NFW form.

Following the introduction of the Navarro-Frenk-White profile (NFW) (Navarro, Frenk \& White 1995,1996,1997), Bartelmann (1996) was the first to consider its lensing implications, showing that the profile is capable of producing radial arcs like those observed in several clusters (e.g. Mellier, Fort \& Kneib 1993; Smail et al. 1995). There have been many subsequent efforts contrasting resultant lensing predictions with observation. Bartelmann et al. (1997) found the optical depth for large arcs for lensing by NFW profiles and found significant dependence on cosmology, though no models could reproduce the observed number. Wright \& Brainerd (2000) and Asano (2000) have contrasted weak lensing mass measurements and 2-image system image positions flux ratios based on the NFW with those for a singular isothermal sphere (SIS). Maoz, Rix, Gal-Yam \& Gould (1997) carried out a survey for wide separation lensed quasars, and compared their results with a thorough statistical study of the number of expected cluster lensed quasars. Modeling clusters using the NFW profile, they found that their null result for observed lensed quasars was consistent with the lensing statistics. Molikawa, Hattori, Kneib \& Yamashita (1999) investigated giant arc statistics and found that the NFW profile can reproduce the number of observed lensed arcs. However their study finds some discrepancy between virial and X-ray temperature for some clusters. Broadhurst, Huang, Frye \& Ellis (2000) have found that an NFW profile (singular core) can explain strong lensing in CL0024+1654, contrary to earlier claims of a flat core (Tyson et al. 1998). However Shapiro \& Iliev (2000) find that the velocity dispersion implied by the NFW fit is a factor of two larger than that observed. The shallower cores resulting from SIDM simulations may resolve this discrepancy. On the other hand, Smith et al. (2000) find that radial arcs imply a core in A383 which has a steeper logarithmic slope than an NFW profile. Williams, Navarro \& Bartelmann (1999) find that arc properties for the most massive clusters are consistent with NFW profiles derived from N-body simulations, but that clusters with velocity dispersions of $\sim1000\,km\,sec^{-1}$ require the presence of a massive central galaxy to reconcile the models with observation. Fox, \& Pen (2000) have considered lensing by galaxy groups modeled as NFW profiles in the Hubble Deep Field. They find that the null result for lenses is consistent with lensing probability. In addition, they compare the probability for multiple imaging due to an NFW profile with the probability for an SIS, and find that the probability is significantly reduced.

Recent complimentary studies have also considered lensing statistics for Zhao profiles. Molikawa \& Hattori (2000) find that the ratio of radial to tangential arcs is a sensitive function of the inner slope of the density profile. Li \& Ostriker (2000) consider a semi-analytic treatment of lensing by the Zhao profile. They find that lensing probability is sensitive to the profile considered, as well as cosmology. In addition, they find a probability for large splittings which is inconsistent with the observed distribution.

\section{GRAVITATIONAL LENSING BY GENERALIZED NFW PROFILES}
\label{lensing}
The NFW halo density profile can be generalized to describe a profile $\rho(r)$ as a function of radius $r$ with an arbitrary power-law shaped central cusp $\rho(r)\sim r^{-\beta}$ and outer regions that fall off as $r^{-3}$ (Zhao 1996),
\begin{equation}
\rho(r)=\frac{\delta_c \rho_c}{\left(\frac{r}{r_s}\right)^{\beta}\left(1+\frac{r}{r_s}\right)^{3-\beta}}
\label{prof}
\end{equation} 
Navarro, Frenk and White (1996) defined the {\em concentration parameter}
\begin{equation}
C_{NFW}\equiv\frac{r_{200}}{r_s},
\end{equation} 
where 
\begin{equation}
r_{200}\equiv\left(\frac{3 M_{200}}{800\pi\rho_c}\right)^{\frac{1}{3}}
\end{equation} 
is the radius enclosing a mass $M_{200}$ with an over-density of 200 above the critical density. The {\em characteristic over-density} $\delta_c$ in Eqn.~\ref{prof} is
\begin{equation}
\delta_c=\frac{200 C_{NFW}^3}{3 F(C_{NFW})}
\end{equation} 
where
\begin{equation}
F(y)\equiv\int_0^y x^{2-\beta}(1+x)^{\beta-3}dx.
\end{equation} 
Therefore, given a cluster mass $M_{200}$, the Zhao profile has two free parameters $\beta$ and $C_{NFW}$. $C_{NFW}$ is quoted for a profile at red-shift zero. Our lensing calculations assume a non-evolving cluster-lens population. The concentration parameter $C_{NFW}(z)$ and characteristic over-density $\delta_c(z)$ of a halo placed at red-shift $z$ are therefore the solution of 
\begin{equation}
C_{NFW}(z)\hspace{5mm}:\hspace{5mm}C_{NFW}^3(z)=\frac{F(C_{NFW}(z))}{F(C_{NFW})}\frac{C_{NFW}^3}{(1+z)^2(1+\Omega_o z)}
\label{cnfwz}
\end{equation}   
and 
\begin{equation}
\delta_c(z)=\frac{\delta_c}{(1+z)^2(1+\Omega_o z)}.
\end{equation}   
These transformations leave $r_s $ and $\rho_c\delta_c$ independent of $z$. From the study of the evolution of a large sample of $\Lambda$CDM halos, Bullock et al. (2000) find $C_{NFW}(\beta=1)\propto(1+z)^{-1}$, consistent with Eqn.~\ref{cnfwz}. Note however, that the inclusion of a finite self-interaction cross-section should result in a different profile evolution.

The Zhao profile is spherically symmetric, while real clusters, and N-body simulations tend to be elliptical or tri-axial. Blandford \& Kochanek (1987) and Kochanek \& Blandford (1987) find that the the introduction of ellipticities smaller than $0.2$ into nearly singular profiles has little effect on the cross-section and image magnification. Furthermore, the images remain on opposite sides of the potential so that the variation in image separations is similar to the axis ratio of the potential. However profiles containing a finite core and ellipticity show a significant increase in cross-section and qualitative differences in image positions. 

We make the thin screen approximation for gravitational lensing by extended bodies. The surface mass density $\Sigma(\xi)$ at a radius $\xi$ may be written
\begin{equation}
\Sigma(\xi)=2\delta_c\rho_c r_s x^{1-\beta} \int_0^{\frac{\pi}{2}} sin\theta (sin\theta+x)^{\beta-3}d\theta,
\label{surf}
\end{equation} 
where $x\equiv\frac{\xi}{r_s}$.
The mass $M(\xi)$ enclosed within the cylinder of radius $\xi$ and the bend angle $\alpha(\xi)$ of a ray with impact parameter $\xi$ are  
\begin{equation}
M(\xi)=2\pi r_s^2\int_0^x x'\Sigma(x') dx'.
\label{mass}
\end{equation} 
and
\begin{equation}
\alpha(\xi)=\frac{4GM(\xi)}{c^2 \xi}.
\label{bend}
\end{equation} 
If $\beta=1$, Eqns.~\ref{surf}, \ref{mass} and \ref{bend} reduce to the expressions computed by Bartelmann (1996).

The lens equation describes the location $\xi$ of an image in the lens plane given a source position $\eta$ (both measured from the observer source line of sight). The angular diameter distances from the observer to a lens at red-shift $z_d$, from the observer to a source at $z_s$ and from the lens to the source are denoted by  $D_d$, $D_s$ and $D_{ds}$. With these, the lens equation may be written (e.g. Schneider, Ehlers \& Falco 1992)
\begin{equation}
0=\theta-\beta-\alpha(\theta)\frac{D_{ds}}{D_s},
\end{equation} 
where $\theta\equiv \frac{\xi}{D_d}$ and $\beta\equiv \frac{\eta}{D_s}$ are the angular positions of the image and source respectively. The point source magnification $\mu_i$ of an image at $\theta_i$ is
\begin{equation}
\mu_i = \frac{\theta_i}{\beta}\left(\frac{1}{1-\frac{D_{ds}}{D_s}\frac{d\alpha(\theta_i)}{d\theta}}\right).
\end{equation} 
The Zhao profile has a critical radius $\eta_{crit}$ separating regions of the source plane that produce 1 or 3 images. The corresponding critical impact parameter $\xi_{crit}$ is the solution of 
\begin{equation}
\xi_{crit}:\frac{4G}{c^2}\left(2\pi\Sigma(\xi)-\frac{M(\xi)}{\xi}\right)-\frac{D_s}{D_d D_{ds}}=0 
\end{equation} 
and the critical angle for multiple imaging is
\begin{equation}
\beta_{crit} = \frac{\xi_{crit}}{D_d}-\alpha(\xi)\frac{D_{ds}}{D_s}.
\end{equation} 
For later comparison, the bend angle for a SIS is 
\begin{equation}
\alpha^{SIS} = \frac{2\pi G M_{200}}{c^2 r_{200}}.
\end{equation}
In this case $\xi_{crit}^{SIS}=0$ so that
\begin{equation}
\beta_{crit}^{SIS} = -\frac{2\pi G M_{200}}{c^2 r_{200}}\frac{D_{ds}}{D_d}.
\end{equation} 

For the computation of lens statistics for an unevolving population of lenses with constant co-moving number density $n_d = n_o(1+z)^3$ we find (Turner, Ostriker \& Gott 1984) that for an $\Omega=1$ universe the differential and total cross-section may be written
\begin{equation}
d\tau(z_d,z_s) = \pi n_o \frac{c}{H_o}D_d^2(z_d)\beta_{crit}^2(z_d,z_s)\sqrt{1+z_d}dz_d
\label{eqn_difod}
\end{equation} 
and
\begin{equation}
\tau(z_s) = \pi n_o\frac{c}{H_o}\int_0^{z_s}D_d^2(z_d)\beta_{crit}^2(z_d,z_s)\sqrt{1+z_d} dz_d.
\label{eqn_od}
\end{equation} 
Turner, Ostriker \& Gott (1984) defined a dimensionless quantity $F_{SIS}$ that measures the efficiency of cosmologically distributed SISs for producing multiple images.
\begin{equation}
F_{SIS}\equiv 4\pi^3 n_o \left(\frac{c}{H_o}\right)^3\left(\frac{G M_{200}}{r_{200}c^2}\right)^2.
\label{F}
\end{equation}
We use $F_{SIS}$ to normalize $\tau$. More realistic calculations of $\tau(z_s)$ for the NFW ($\beta=1$) profile have previously been made within the contexts of wide separation quasars (Maoz, Rix, Gal-Yam \& Gould 1997), giant arc statistics (Bartelmann et al. 1997; Molikawa, Hattori, Kneib \& Yamashita 1999) and distant supernovae (Porciani \& Madau 2000). However in this work we are interested in relative values of $\tau$ produced by different profiles, and so have neglected quantitatively important effects such as formation history.    

One image is present for all source positions. We denote this image to be at $\theta_1$. Note that $\theta_1>\beta_{crit}$ for all $\beta$. For $\beta<\beta_{crit}$ there are two additional images with positions $\theta_2<\beta_{crit}$ and $\theta_3>\beta_{crit}$. We label the corresponding point source image magnifications as $\mu_1$, $\mu_2$ and $\mu_3$ respectively. If $\beta_{crit}>0$ then $\theta_1>0$, $\theta_1,\theta_2<0$, $\mu_1,\mu_3>0$ and $\mu_2<0$. In addition $|\mu_3|\ge|\mu_2|$. For $\beta<\beta_{crit}$,  we define the average image separation of a multipli-imaged source at $z_s$ due to a deflector at $z_d$ to be 
\begin{equation}
\langle\Delta\theta(z_d,z_s)\rangle = \frac{2}{\beta_{crit}^2}\int_0^{\beta_{crit}}\beta(\theta_1(\beta,z_d,z_s)-\theta_3(\beta,z_d,z_s))d\beta.
\end{equation} 
Similarly the average magnification of a multipli-imaged source at $z_s$ due to a deflector at $z_d$ is 
\begin{equation}
\langle\mu(z_d,z_s)\rangle = \frac{2}{\beta_{crit}^2}\int_0^{\beta_{crit}}\beta(|\mu_1(\beta,z_d,z_s)|+|\mu_2(\beta,z_d,z_s)|+|\mu_3(\beta,z_d,z_s)|)d\beta.
\end{equation} 
It follows that the average image splittings $\langle \Delta\theta(z_s)\rangle$ and total magnifications $\langle \mu(z_s)\rangle$ for a source at $z_s$ are
\begin{equation}
\langle \Delta\theta(z_s)\rangle=\frac{1}{\tau(z_s)}\int_0^{z_s}\langle\Delta\theta(z_d,z_s)\rangle\frac{d\tau(z_d,z_s)}{d z_d}d z_d
\label{sep}
\end{equation} 
and 
\begin{equation}
\langle \mu(z_s)\rangle=\frac{1}{\tau(z_s)}\int_0^{z_s}\langle\mu(z_d,z_s)\rangle\frac{d\tau(z_d,z_s)}{d z_d}d z_d.
\label{mag}
\end{equation} 

\section{CROSS-SECTION, IMAGE SPLITTINGS AND MAGNIFICATION}
\label{results}

In this section we present the differential and total cross-sections (Eqns.~\ref{eqn_difod} and \ref{eqn_od}), distributions of image separations, and distributions of magnifications (Eqns.~\ref{sep} and \ref{mag}) for Zhao profiles having different parameters $\beta$ and $C_{NFW}$.

Fig.~\ref{profsurf} shows $\rho(r)$ (top) and $\Sigma(\xi)$ (bottom) for three different Zhao profiles. These are labeled profile $I$ (solid lines), $II$ (dashed lines) and $III$ (dotted lines), have parameters of $(\beta,C_{NFW})=(1.5,20)$, $(1.0,15)$, and $(0.5,10)$ respectively, and are increasingly steep cusps in their central regions in addition to being increasingly centrally concentrated. The model clusters have an $M_{200}$ equal to that of a SIS ($M_{200}^{SIS}$) with a velocity dispersion corresponding to a large cluster $\sigma_{SIS}^{cstr}=1000\,km\,sec^{-1}$:     
\begin{equation}
M_{200}^{SIS}=(\sigma_{SIS}^{cstr})^3\left(\frac{3}{800\pi\rho_c}\right)^{\frac{1}{2}}\left(\frac{2}{G}\right)^{\frac{1}{3}}.
\end{equation}
The solid light lines show $\rho_{SIS}(r)$ and $\Sigma_{SIS}(\xi)$ of the SIS for comparison. The horizontal dashed lines show the critical surface mass density 
\begin{equation}
\Sigma_{crit} = \frac{c^2}{4\pi G}\frac{D_s}{D_d D_{ds}}
\end{equation}
for a source at $z_s=3$ and lenses at $z_d=0.5$, $1.0$ and $1.5$. Comparison of $\Sigma_{crit}$ with $\Sigma(\xi)$ indicates a significant variation of the lensing probability between profiles having the same $M_{200}$ but different profile parameters. Profiles $I$ and $II$ reach the critical density over a much larger area than profile $III$, which is only just supercritical at $z_d=1.0$, and sub-critical at $z_d=2.0$. 

The variation of cross-section with the profile parameters is quantified by Fig.~\ref{od} which shows $\tau(z_s)$ for Zhao profiles $I-III$. The cross-section is given in units of $F_{SIS}$ (Eqn.~\ref{F}). For comparison, the cross-section for a SIS ($\tau_{SIS}(z_s)$, light line, Turner, Ostriker \& Gott 1984) is also shown in Fig.~\ref{od}. Examples of the effect of different cosmologies on $\tau_{SIS}(z_s)$ (and on $\frac{1}{\tau(z_s)}\frac{d\tau(z_s,z_d)}{d z_d}$ in Fig.~\ref{difod}) can be found in Turner et al. (1984) and Turner (1990). There is a striking variation in $\tau$ over these three profiles. Profiles $II$ and $III$ produce $\tau$s which differ by an order of magnitude, while there are 4 orders of magnitude between $\tau$ for profiles $I$ and $III$. More concentrated profiles are much more efficient for producing multiple images due to their higher central densities. For larger source red-shifts the SIS has a similar cross-section to profile $II$, while at smaller red-shifts the cross-section is similar to that of profile $I$, indicating  that different profiles are  efficient lenses at different red-shifts. 

In Fig.~\ref{difod} we have plotted $\frac{1}{\tau(z_s)}\frac{d\tau(z_s,z_d)}{d z_d}$ for profiles $I$ (solid lines), $II$ (dashed lines) and $III$ (dotted lines) with $z_s=1.5$ (upper panel) and $z_s=3.0$ (lower panel). The light lines show the corresponding distributions for a SIS. Less concentrated profiles have distributions that are more sharply peaked, with modes at lower red-shift (c.f. Kochanek \& Blandford~1987). The lower cross-section to multiple imaging is therefore due to a combination two effects. Firstly, the halo reaches the critical density to lensing only over a small region in the center due to the low concentration or central slope. In addition, there is a decreased redshift range in which the central surface density exceeds the critical value anywhere. Note that the assumption in Eqns.~\ref{eqn_difod} and \ref{eqn_od} of an unevolving population of SIDM halos is aided in the case of low-concentration profiles since these preferentially lens at low red-shift.

To further investigate the variation of cross-section with $\beta$ and $C_{NFW}$ we have plotted contours of $\frac{\tau(z_s)}{\tau_{SIS}(z_s)}$ with $z_s=3$, over a parameter space of $\beta$ and $C_{NFW}$ (Fig.~\ref{contod}). Two additional dark lines are shown. The upper dark line is a contour of unity and marks the parameters that produce an cross-section equal to that of a SIS. The lower dark line marks the boundary of the region of parameter space where no multiple imaging is produced (on this and subsequent plots this line is the approximated by the $10^{-5}$ contour). The contours describe sets of parameters $\beta$ and $C_{NFW}$ that are equally efficient for producing multiple images. Profiles $I$, $II$ and $III$ lie on a line in $\beta-C_{NFW}$ space that is approximately orthogonal to the contours of $\tau(z_s)$. However, figure~\ref{contod} demonstrates that dramatic variability in $\tau(z_s)$ may be obtained by variation of either $\beta$ or $C_{NFW}$. 

We have calculated $\frac{\tau(z_s)}{\tau_{SIS}(z_s)}$, $\langle\Delta\theta(z_s)\rangle$ and $\langle \mu(z_s)\rangle$ over a small grid of $\beta$ and $C_{NFW}$ (see Tab.~\ref{sepmagtab}). We find that profiles which produce a small cross-section to multiple imaging also produce smaller image splittings, although only by a factor of a few compared with more highly concentrated profiles. However, the image magnification is significantly larger for the less concentrated profiles. Distributions of image separations ($\frac{dP}{d\Delta\theta(z_s)}$) for the Zhao profile are presented in Fig.~\ref{sep_dist} for the values of $\beta$ and $C_{NFW}$ in Tab.~\ref{sepmagtab}. For comparison, the light line shows the distribution of splitting angles for a SIS ($\frac{dP_{SIS}}{d\Delta\theta(z_s)}$). There is a trend for the distributions with lower $\langle\Delta\theta(z_s)\rangle$ to also have a narrower range of $\Delta\theta(z_s)$. In contrast to the SIS distribution, the Zhao profile distributions are asymmetric about the mode, having a longer large splitting tail. Fig.~\ref{amp_dist} shows magnification distributions $\frac{dP}{d\mu}$ for the values of $\beta$ and $C_{NFW}$ in Tab.~\ref{sepmagtab}. The distribution for an SIS, 
\begin{equation}
\frac{dP^{SIS}}{d\mu}=\frac{8}{\mu^3},
\label{PSIS}
\end{equation}
is shown for comparison (light line). From Fig.~\ref{amp_dist} we see that the high mean magnifications shown in Tab.~\ref{sepmagtab} result from a higher minimum magnification of multiple images. Lenses with low concentration and small $\beta$ result in a low $\tau(z_s)$. Very close alignment is therefore required to produce multiple imaging, and the close alignment results in high magnification. More concentrated lenses do not require such high alignment for multiple imaging. This results in a larger cross-section and reduced mean magnification.

The basic lensing properties of the Zhao profile may therefore be summarized as follows. Profiles that have a small concentration and flat central profile produce significantly less multiple imaging than a profile with either a high concentration or steep central profile (or both). However, these few multiple images are significantly magnified. In any lens survey, the increased $\langle \mu \rangle$ will partially offset the decreased $\tau(z_s)$ in determining the lens frequency. This magnification bias is discussed in Sec.~\ref{amp_dist_bias}. For clusters of galaxies, the reduced splittings for low density profiles are $\ga10''$ and should not pose a significant selection bias for well defined surveys.

\section{THE $C_{NFW}-\beta$ DEGENERACY}
\label{degeneracy}

A property of the Zhao profile (Eqn.~\ref{prof}) is the ability to describe a high central density either by a steep central cusp, or by a large concentration parameter (so that the turnover separating the core regions from the outer halo occurs closer to the center). This degeneracy is present between other profiles exhibiting a power-law core and a characteristic radius. Klypin, Kravtsov, Bullock \& Primack (2000) have investigated the relative suitability of Moore et al. (1999) and NFW ($\beta=1$) profiles as descriptions of CDM halos. For galaxy size halos, they find differences which are small at radii above $0.01 r_{200}$. Larger differences are found for lower concentration cluster halos as the profiles central slope begins to approach its asymptotic value. Moreover, using a very general 4-parameter NFW profile, Klypin et al. (2000) obtained equally good fits to an N-body halo for several different sets of parameters. Our interest lies in the relationship between the $C_{NFW}-\beta$ degeneracy found in in the contexts of multiple imaging statistics and profile fits to simulated halos.

Our approach is to select a collection of profiles $\rho_o(r)$ having $\beta=0.5$ but different concentrations $C_{NFW,o}$. Keeping $M_{200}$ constant (for comparison with our strong lensing calculations), we fit a new profile $\rho_{fit}(r)$ with $C_{NFW}$ as a function of $\beta$. Fits are obtained by minimizing each of three different quantities which emphasize different parts of the halo:
\begin{equation}
\chi^{2}=\int_{\frac{r_{200}}{100}}^{r_{200}} r^2(\rho(r)-\rho_o(r))^2 dr,
\end{equation}
\begin{equation}
\Delta\rho_{max}=max(|log(\rho(r))-\log(\rho_o(r))|)
\end{equation}
(Klypin, Kravtsov, Bullock \& Primack 2000), and 
\begin{equation}
\chi_{rel}^{2}=\int_{\frac{r_{200}}{100}}^{r_{200}} r^2\left(\frac{\rho(r)-\rho_o(r)}{\rho_o(r)}\right)^2 dr.
\end{equation}
Minimization is performed over the range $0.01r_{200}-r_{200}$. 

Some examples of the fitted profiles and their fractional residuals are presented in Fig.~\ref{fitprofa} (light lines). The dark lines show the profile being fitted. In each case we quote the parameters $\beta_o$, $C_{NFW,o}$ and $\beta$, $C_{NFW}$, as well as the average absolute relative difference over the fitting volume
\begin{equation}
\langle \frac{\Delta \rho}{\rho}\rangle= \frac{3}{r_{200}^3-\left(\frac{r_{200}}{100}\right)^3} \int_{\frac{r_{200}}{100}}^{r_{200}}r^2\frac{|\rho_{fit}(r)-\rho_o(r)|}{\rho_o(r)}dr.
\end{equation}
The central density dominates minimization of $\chi^{2}$ (top panels) so that the fitted profile has the correct central density at the cost of error over large regions around $r=\frac{r_{200}}{10}$. Alternatively if we set $C_{NFW}$, and adjust $\beta$, the fitted profile has the correct central slope and density, but is systematically high or low over the outer regions. The initial and fitted halos have average absolute relative differences of a few to $\sim15\%$. As noted by Klypin, Kravtsov, Bullock \& Primack (2000), minimization of $\Delta\rho_{max}$ (central panels) often does a much better job of reproducing the density at small $r$. If $\beta$ is the fitted parameter rather than $C_{NFW}$, a similar result is obtained, the central slope adjusts so as to give the correct density at small $r$. Minimization of $\Delta\rho_{max}$ produces halo fits with similar average absolute relative differences to $\chi^{2}$. Finally, minimizing $\chi_{rel}^{2}$ (lower panels) obtains good fits beyond $r\sim0.1r_{200}$ at the expense of an incorrect central density. The mean relative error is substantially reduced, with average absolute relative differences of a fraction of a percent.

 The dot-dashed lines in the upper, central and lower plots of Fig.~\ref{fitprofb} are depictions of the $\beta-C_{NFW}$ degeneracies corresponding to the minimization quantities $\chi^{2}$, $\Delta\rho_{max}$ and $\chi_{rel}^{2}$. Since $\chi^{2}$ produces good fits to the core region, which is known to be most important for strong lensing (e.g. Turner, Ostriker \& Gott 1984), the lines of degeneracy are nearly parallel to those describing the cross-section to multiple imaging over much of the parameter space. However for combinations of low or high concentration and low inner slope, the approximately degenerate profiles do not provide a accurate description of the lensing statistics. Minimization of $\Delta\rho_{max}$ results in profile degeneracies similar to those for strong lensing unless the profile has large $C_{NFW}$ and small $\beta$. Finally, minimization of $\chi_{rel}^{2}$ leads to degeneracies that differ significantly from those for strong lensing. Fits that are weighted towards the central regions, and which result in highly concentrated profiles may therefore be converted to the analytically treatable NFW ($\beta=1$) profile while approximately preserving the lensing statistics. This is particularly true for fits based on minimization of $\Delta\rho_{max}$. On the other hand, lensing calculations based on profiles which best describe the density over the largest volume (out to $r_{200}$) may severely under or over-estimate the strong lensing statistics. For example, Zhao halos having the same $M_{200}$, but different parameters ($\beta,C_{NFW}$) = (1.5,6.0) and (1.0,10.0) are degenerate to fitting by the minimization of $\chi_{rel}^{2}$, but differ in $\tau$ by an order of magnitude. Thus Fig.~\ref{fitprofb} may be considered a quantification (for Zhao profiles) of the statement that lensing is sensitive to the size of the halo core.

\section{MAGNIFICATION BIAS}
\label{amp_dist_bias}

Statistics generated from surveys for gravitational lenses are subject to several classes of bias. Kockanek (1991) has given a detailed discussion of selection biases in optical, imaging, galaxy mass gravitational lens surveys. Two primary sources of bias against finding a lens are selection against small separations due to the finite angular resolution of the survey, and the obscuration or differential reddening of one or both images due to dust in the lens galaxy. When considering lensing by clusters, the splittings are nearly always larger than the survey resolution. In addition, the lensed images are viewed through the halo rather than the outskirts of a galaxy. These biases should not therefore operate on any significant level in the analysis of cluster lensing statistics. The magnification of a lensed image has the potential to boost the observed flux of a source which is intrinsically faint above the survey detection limit. Lensed images therefore sample fainter members of the source population. This is magnification bias (Gott \& Gunn 1974; Turner 1980), and it a significant factor in any lens survey.

As mentioned in Sec.~\ref{results}, lenses with low concentration and a shallow cusp result in a low cross-section. Very close alignment is therefore required to produce multiple imaging, and this results in a high minimum magnification. More concentrated lenses do not require such high alignment to produce multiple imaging, thus the cross section is larger, and the minimum magnification not so high. In the latter case, although the mean is lower, high magnifications are still present where the alignment is high. As a result, the mean magnification is not necessarily a fair indication of the magnification bias. 

Magnification bias operates differently in surveys for galaxy and cluster lenses. The typical separation of a galaxy lens is smaller than most optical survey resolution limits, and the total magnification must be used to compute the magnification bias. However, the total magnification is not the important quantity for cluster lensing surveys since the separate images are generally  resolved (see also Chiba \& Yoshii 1997). There are different approaches to making a wide separation gravitational lens survey. Firstly, candidate lenses may be selected from galaxy or cluster catalogues having one image of a source in good alignment with a potential lens. Follow up imaging can then be used to look for a second, fainter image. This is the approach taken by systematic surveys to date (e.g. Maoz et al. 1997; Phillips, Browne \& Wilkinson 2000). Here the magnification bias which results in the inclusion of the brighter image into the initial survey is important (though the overall bias for the survey must also consider the detectability of the faint image in follow up observations). 

A second method looks for pairs of lensed images in existing data. In this instance, the magnification bias which also includes the fainter image (the second brightest for three image lenses, i.e. image 3), and thus both images into the survey is the relevant quantity.

To compute magnification bias, a number magnitude relation for the sources is required. We use the function described by Kochanek (1996), based on data from Boyle, Shanks \& Peterson (1988) and Hartwick \& Schade (1990). He uses a broken-power-law form (Turner 1990). To simplify our calculations and to keep the conclusions as general as possible, we assume that the survey depth is deeper than the break B-magnitude at $z_s=3$ of $m_B\sim19$. The magnitude-number relation may then be written as 
\begin{equation}
\frac{dN}{dm}=N_o10^{\alpha_m m}
\label{numbercount}
\end{equation}
where $N_o$ is a normalizing constant and $\alpha_m$ is the logarithmic slope. For a survey with a limit $m_{lim}$, the magnification bias is defined as (Fukugita \& Turner 1991)
\begin{equation}
\label{biaseqn}
B\equiv\frac{\int_0^{\infty}\frac{dP}{d\mu}N\left(<m_{lim}+\frac{5}{2}log_{10}(\mu)\right)d\mu}{N(<m_{lim})}.
\end{equation}
In application to a real lens survey, the magnification bias may be written more conveniently as
\begin{equation}
B\equiv\frac{\int_0^{\infty}\int_{\mu_{min}}^{\mu_{max}}d\mu_b\frac{dP}{d\mu_b}N\left(<m_{lim}+\frac{5}{2}log_{10}(\mu_b)\right)d\mu_b}{N(<m_{lim})},
\end{equation}
where the limits $\mu_{min}$ and $\mu_{max}$ are the range of the bright image magnification $\mu_b$ such that at least one other image is visible. Note that only the magnification distribution for the bright image is directly relevant. In this paper we adopt the simpler and non-specific case of magnification bias for bright and faint images based on Eqn.~\ref{biaseqn}.

Fig.~\ref{amp_dists_fig} shows the magnification distributions for the bright (left hand panel) and faint (right hand panel) images for Zhao profiles. The light lines show the distributions for a SIS:
\begin{equation}
\frac{dP^{SIS}}{d\mu_{f}}=\frac{2}{(\mu_{f}+1)^3}\:\:for\:\mu_{f}>0
\label{PSIS_faint}
\end{equation}
and 
\begin{equation}
\frac{dP^{SIS}}{d\mu_{b}}=\frac{2}{(\mu_{b}-1)^3}\:\:for\:\mu_{b}>2
\label{PSIS_bright}
\end{equation}
for comparison. The image magnification distributions for the Zhao profiles exhibit similar behavior to the SIS, and there is a high probability that the two images will have different magnifications. However unlike the SIS, the Zhao profiles have finite $\frac{d\alpha(\theta)}{d\theta}$ as $\theta\rightarrow 0$. The faint image therefore has a finite minimum magnification. 

The magnification bias described by Eqn.~\ref{biaseqn} describes the additional number of lensed pairs. If the quantity of interest is the density of lensed images on the sky, then an additional factor of $\mu^{-1}$ must be inserted under the integral to account for the smaller region of source plane sampled per unit area of lens plane. Inserting Eqn.~\ref{numbercount} into Eqn.~\ref{biaseqn} we find 
\begin{equation}
B_{b}=\int_0^{\infty}\frac{dP}{d\mu_{b}} \mu_{b}^{\frac{5}{2}\alpha_m}d\mu_{b}
\label{Bbright}
\end{equation}
for magnification bias due to bright images, and
\begin{equation}
B_{f}=\int_0^{\infty}\frac{dP}{d\mu_{f}} \mu_{f}^{\frac{5}{2}\alpha_m}d\mu_{f}
\label{Bfaint}
\end{equation}
for magnification bias due to faint images. Note that the magnification bias is equal to the mean magnification for $\alpha_m=0.4$. Since the high magnification tail of any magnification distribution has 
\begin{equation}
\frac{dP}{d\mu}\propto\mu^{-3},
\end{equation}  
the integral in Eqns.~\ref{Bbright} and \ref{Bfaint} is finite for $\alpha_m<\frac{4}{5}$, and in this range no low-magnitude cutoff need be assumed.

Using Eqns.~\ref{PSIS_faint} and \ref{PSIS_bright} the magnification biases for the faint and bright images of the SIS are 
\begin{equation}
B^{SIS}_{b}=\int_2^{\infty}\frac{2}{(\mu_{b}-1)^3}\mu_{b}^{\frac{5}{2}\alpha_m}d\mu_{b}.
\end{equation}
and
\begin{equation}
B^{SIS}_{f}=\int_0^{\infty}\frac{2}{(\mu_{f}+1)^3}\mu_{f}^{\frac{5}{2}\alpha_m}d\mu_{f}.
\end{equation}

Fig.~\ref{bright_amp_bias} shows the absolute magnification bias (left hand panel), and the magnification bias relative to the SIS (right hand panel) based on the bright image as a function of $\alpha_m$. Plots are shown for the values of $\beta$ and $C_{NFW}$ in Tab.~\ref{sepmagtab}. The light line in the left hand panel shows the magnification bias curve for the SIS. Fig.~\ref{faint_amp_bias} shows the corresponding results for the faint image. Values of absolute and relative (to the SIS) magnification bias based on the faint and bright image are presented in Tabs.~\ref{abbright}, \ref{abbrightSIS}, \ref{abfaint} and \ref{abfaintSIS} for $\alpha_m=0.2,0.4,0.6$. For quasars Kockanek (1996) finds $0.27\pm0.07$. 

The absolute magnification biases for the bright images range up to a few, a few tens and a few hundred for $\alpha_m=0.2,0.4$ and $0.6$ respectively. However, as discussed at the top of this section, the quantity of interest is the relative magnification bias. The values in Tab.~\ref{abbrightSIS} demonstrate the reduced amplitude of the relative (to the SIS) bias, which is similar for $\alpha_m=0.2$, but reduced by a factors of $\sim3$ and $\sim5$ for $\alpha_m=0.4$ and $0.6$. Similar behavior is seen in magnification bias of faint images. The absolute magnification biases for the faint images produced by Zhao profiles are slightly reduced from those computed for the bright image. However the magnification biases for the faint images relative to the SIS are increased by a factor of two over the corresponding bright image case. This apparent contradiction arises because the faint image produced by a SIS may be arbitrarily faint. 

The magnification bias results in additional sources being available for lensing. Since the relative magnification bias is greater than 1 where the ratio of cross-sections is less than one, the magnification bias counteracts the dramatic dependence of $\tau$ on $\beta$ and $C_{NFW}$ seen in Fig.~\ref{contod}. The relevant quantity is therefore the optical depth to multiple imaging $\tau(z_s) B_{b}$ or $\tau(z_s) B_{f}$. Tabs.~\ref{abbrighttao} and \ref{abfainttao} show values of the optical depth relative to the SIS for biases computed using bright and faint images. Values are shown assuming $\alpha_m=0.2$, $0.4$ and $0.6$. The magnification bias reduces the difference between cross-sections for different profiles, though the variation with $\beta$ and $C_{NFW}$ is still very dramatic. However, profiles with no cross-section to multiple imaging are not subject to magnification bias. Magnification bias therefore increases the dependence of the cross-section to multiple imaging on $\beta$ and $C_{NFW}$ near the multiple-imaging / no-multiple-imaging boundary.

\section{IMPLICATIONS FOR CDM AND SIDM PROFILES}
\label{implications}

Miralda-Escude (2000) argued that strong lensing was a powerful test of SIDM models. Assuming a cross-section independent of velocity, he pointed out that if SIDM were responsible for the cores in dwarf galaxies, then scaling arguments lead to clusters cores that are larger than observed. While Miralda-Escude used a scaling law that is not a good fit to the SIDM simulations (Dav$\acute{\mathrm{e}}$, Spergel, Steinhardt \& Wandelt 2000), his basic argument, that lensing is a powerful test of the model remains valid.  Furthermore, Miralda-Escude~(2000) points out that the mis-alignment of the images in the cluster lens MS2137-23 cannot be reconciled with a spherical lens implied by SIDM halos. 

Recently, Yoshida, Springel, White \& Tormen (2000) have produced simulations of a cluster with $M_{200}=7.4\times 10^{14}M_{\odot}$. They model the cluster using collision-less cold dark matter (labeled S1), as well as weakly self interacting cold dark matter having collision cross-sections of $\sigma^{\star}=0.1$, 1.0 and $10.0cm^2gm^{-1}$ (labeled S1W-a, S1W-b, and S1W-c). Halo density profiles are presented at $z=0$ for all 4 realizations. In addition, they present halo profiles for the simulation S1W-c at a range of different epochs. During the evolution of the halo, the central densities are most extreme when the cosmological expansion factor $a$ (normalized to its present value) takes the values $a=0.78$ and $a=1.72$. The 6 halo profiles are reproduced in Fig.~\ref{yoshfit} (dots). To estimate the lensing rate due to these different profiles we have fitted each with the Zhao profile (Eqn.~\ref{prof}). The fitting was achieved via minimization of $\chi^2$ (solid lines) and $\Delta\rho_{max}$ (dashed lines). The constraint that $\beta\ge0.2$ was imposed and the halos S1W-c ($a$=0.78) and S1W-c ($a$=1.72) were fitted assuming an $r_{200}$ (and hence $M_{200}$) corresponding to that at $a$=1. Yoshida et al. (2000) note that the halo approximately doubled its $M_{200}$ between $a=0.73$ and $a=1.72$. If half this gain was made before (after) $a=1$, then $C_{NFW}$ for the halo S1W-c ($a$=0.78) (S1W-c ($a$=1.72)) should be revised downwards (upwards) by $\sim10\%$. Both the fits and fractional residuals are plotted in Fig.~\ref{yoshfit}. The Zhao profile provides an adequate fit to halos S1 and S1W-a. However for the larger interaction cross-sections, the core density is overestimated by the fitted profile. The corresponding lensing rate may therefore also be overestimated in these cases.

The estimates for the profile parameters $C_{NFW}$ and $\beta$ of the Yoshida, Springel, White \& Tormen (2000) halos are plotted over contours of $\frac{\tau(z_s)}{\tau_{SIS}(z_s)}$ in Fig.~\ref{yosh_points}. Each parameter set is labeled and the upper and lower plots show points corresponding to the fits obtained by minimization of $\chi^2$ and $\Delta\rho_{max}$ respectively. From Fig.~\ref{yosh_points} we find that $\tau$ falls off very rapidly for small $\beta$ and $C_{NFW}$, the region of interest for SIDM halos. The halos S1W-b and S1W-c at the current epoch do not produce any multiple images. The halo S1W-a (at the current epoch) lenses at a significantly reduced rate with respect to the CDM halo S1. The variation in the density profile at different epochs is due to successive mergers (Yoshida et al. 2000). The spread in parameters obtained from profile fitting is representative of the range of lensing strengths expected. A population of S1W-c ($a=1.72$) halos would yield a finite lensing rate, comparable with that of S1W-a halos, while at the other epochs considered ($a=0.78$ and $a=1.0$) the halos would produce no multiple images.

For comparison, we have also included in Fig.~\ref{yosh_points}, points corresponding to parameters found by two complimentary studies of CDM halos. Jing \& Suto (2000) have estimated of the spread in the distribution of concentration parameter for CDM halos by fitting the Zhao form to 4 cluster mass halos ($M_{200}\sim3-5\times 10^{14}M_{\odot}$) using both $\beta=1$ and $\beta=1.5$. Their results are plotted in the lower panel of Fig.~\ref{yosh_points} (labeled 1-4 as per Jing \& Suto (2000)). This range is in agreement with the 68\% range for $C_{NFW}$ (extrapolated to $M_{200}\sim10^{15}M_{\odot}$) found by Bullock et al. (2000) (plotted in the upper panel of Fig.~\ref{yosh_points}). The halo S1 has a $C_{NFW}$ which is consistent with the range of values found by these authors. It is important to note that the range of parameters describe halos that produce lensing cross-sections differing by several orders of magnitude. The lensing statistics will therefore be dominated by the more concentrated halos, indicating that the medians of profile parameter distributions may not provide a reasonable estimation of the lensing rate for a particular dark matter model. We return to this point in the following section.

More detailed modeling including evolution, non-spherical mass distributions, and a range of formation red-shifts, as well as observational details including magnification bias etc. (e.g. Turner, Ostriker \& Gott 1984; Kochanek 1991) appropriate for surveys such as that being undertaken by the SDSS collaboration (e.g. York et al. 2000), is required before quantitative constraints on SIDM models can be derived. In addition, since lensing probes the central regions of the cluster, the contribution to the potential of a centrally dominant galaxy may be significant and must also be considered. We offer a preliminary discussion of the contribution of central galaxies in Sec.~\ref{galaxies}, however, we conclude that the sensitivity of $\tau$ to $\beta$ and $C_{NFW}$ indicates that cluster lensing statistics will provide a powerful probe of the self-interaction cross-section. 

While lensing constrains the dark matter properties for dark matter collisions 
with velocities of $v\sim10^3\,km\,sec^{-1}$, observations of dwarfs constrain dark matter properties with $v\sim30\,km\,sec^{-1}$. If the dark matter particle is a composite particle, analogous to a nucleon with a low energy resonance, then its cross-section scales as $\sigma^{\star} v = const.$ For fixed cross-sections on the dwarf scale, a cross-section that is inversely proportional to the characteristic velocity will produce significantly smaller, less spherical cores in clusters and large cores in dwarfs (Yoshida, Springel, White \& Tormen  2000), consistent with observations of cluster cores like CL0024+1654 and CL0016+16 (Avila-Reese, Firmani, D'Onghia \& Hernandez 2000; Firmani et al. 2000). Alternatively, if the dark matter can self annihilate at late times (Kaplinghat, Knox \& Turner 2000) then $\sigma^{\star} v=const.$, this leads to low density tri-axial cores and differing lensing predictions.

\section{DEPENDENCE OF OPTICAL DEPTH ON THE HALO PROFILE DISTRIBUTION}
\label{halodist}

Recent studies of N-body simulations have described the distribution of halo concentrations (Bullock et al. 2000; Jing \& Suto 2000;  Dav$\acute{\mathrm{e}}$, Spergel, Steinhardt \& Wandelt 2000). Moreover, the simulations of Yoshida, Springel, White \& Tormen (2000) demonstrate large variation in the concentration of a SIDM halo during its evolution. An important quantity which has hitherto been neglected in parametric cluster lens analyses is the distribution of halo profiles. 

The upper panels of Fig.~\ref{od_spread} show cross-sections $\frac{\tau(z_s)}{\tau_{SIS}(z_s)}$ (thick light lines) and optical depths $\frac{B_f}{B_f^{SIS}}\frac{\tau(z_s)}{\tau_{SIS}(z_s)}$ (thick dark lines) to multiple imaging with respect to an SIS as a function of $C_{NFW}$ for $\beta=1.0$ (left hand panel) and $\beta=0.5$ (right hand panel). We have assumed $\alpha_m=0.2$, appropriate for the faint end of the quasar luminosity function. The light curves are sections through Fig.~\ref{contod}. The corresponding curves for $\frac{B_f}{B_f^{SIS}}\frac{\tau(z_s)}{\tau_{SIS}(z_s)}$ show the decreased slope of $\frac{B_f}{B_f^{SIS}}\frac{\tau(z_s)}{\tau_{SIS}(z_s)}$ with $C_{NFW}$ for large $C_{NFW}$, and the increased slope near the no-multiple-imaging cutoff discussed at the end of Sec.~\ref{amp_dist_bias}. The upper panels also show hypothetical $C_{NFW}$ distributions having means of $\langle C_{NFW}\rangle=6.1$ (thin solid lines) and $\langle C_{NFW}\rangle=10.9$ (thin dashed lines). In each case one delta function and one Gaussian distribution (half-width of $\sigma_{C_{NFW}}=2$) are plotted. The lower panels show the dependence on $\sigma_{C_{NFW}}$ of convolutions of $\frac{\tau(z_s)}{\tau_{SIS}(z_s)}$ (light lines) and $\frac{B_f}{B_f^{SIS}}\frac{\tau(z_s)}{\tau_{SIS}(z_s)}$ (dark lines) with Gaussian distributions of $C_{NFW}$. The solid and dashed lines correspond to convolutions with $C_{NFW}$ distributions having $\langle C_{NFW}\rangle=6.1$ and $10.9$ respectively. The convolutions in the lower left and right panels assumed $\beta=1.0$ and $\beta=0.5$. 

The optical depth shows significant dependence on the range of $C_{NFW}$. For example, the simulations of Bullock et al. (2000) yield $\beta=1$, $\langle C_{NFW}\rangle\sim 6$ and $\sigma_{C_{NFW}}\sim 2$ for CDM halos. The spread of $\sigma_{C_{NFW}}\sim 2$ results in a 2 fold increase in optical depth to multiple imaging. Much more dramatic examples are obtained where $\langle C_{NFW}\rangle$ lies close to, or below the no-multiple-imaging boundary, and the significant portion of optical depth is produced by outliers in the concentration population. This is the region of parameter space important for SIDM halos. We use the S1W-b halo of Yoshida, Springel, White \& Tormen (2000) as an example. Using the $C_{NFW}-\beta$ degeneracy lines in Fig.~\ref{fitprofb}, we find that this halo may be approximated by a Zhao profile with $\beta=0.5$ and $C_{NFW}\sim6$. These parameters place the S1W-b halo in the no-multiple imaging regime. However, if halos have concentrations distributed about $C_{NFW}\sim6.1$, the optical depth is finite. Taking the example of $\langle C_{NFW}\rangle=6$, we find that 2 orders of magnitude separate the optical depth due to distributions of halos with $\sigma_{C_{NFW}}=1$ and $\sigma_{C_{NFW}}=2$, and that 3 orders of magnitude separate the optical depth between distributions having $\sigma_{C_{NFW}}=1$ and $\sigma_{C_{NFW}}=4$. Note that the S1W-c halo of Yoshida et al. (2000) has a concentration that differs by up to $\Delta C_{NFW}\sim8$ from the present day value during its evolution. These results demonstrate that knowledge of the distribution of halo concentrations is an important consideration for cluster lensing studies.

\section{THE CONTRIBUTION OF CLUSTER GALAXIES}
\label{galaxies}

The statistics of multiple imaging due to clusters of galaxies are sensitive to the inner most regions of the mass distribution, where the contribution of a central galaxy may be important. One expects that this is particularly true where the halo is less concentrated or or has a shallower cusp, as is the case for SIDM halos. The presence of an isothermal central galaxy renders the cross-section to multiple imaging finite for all halos.

Williams, Navarro \& Bartelmann (1999) have investigated the effect of including a massive central galaxy. They find that arc properties of the most massive clusters are consistent with predictions of the NFW profile, but that observed separations are too large in the lower mass cluster lenses. They demonstrate that a centrally dominant galaxy with a dark halo can resolve the discrepancy. However there are several caveats. As noted by Williams et al.~(1999), their cluster sample over represents the expected fraction of massive clusters (based on the algorithm of Press \& Schechter 1974), possibly indicating a selection bias towards large splittings and highly elongated arcs. Furthermore, the higher resolution of simulations of Moore et al. (1998) show a steep core $\rho(r)\sim r^{-\frac{3}{2}}$, indicating that the concentrations found by NFW may be a factor of $\sim 2$ too low. This reduces the required mass of the centrally dominant galaxy. More importantly, Bullock et al. (2000) find significant scatter in concentration for a halo of given mass. High concentration halos are more likely to lens, produce larger image splittings, and so are more likely to be observed as cluster lenses. In the previous section this was shown to be a significant effect and might give the appearance that the profiles found from N-body simulations cannot account for observations of cluster lenses. On the other hand, Bullock et al. (2000) also find a concentration that decreases faster with redshift than originally thought (Navarro, Frenk \& White 1997). This will exasperate the problem pointed out by Williams et al. (1999) for cluster lenses above a redshift of $\sim 1$. 

On the other hand, a cluster need not have a single centrally dominant galaxy (e.g. CL0024+1654). Flores, Maller \& Primack (2000) have analyzed the inclusion of cluster galaxies on the statistics of giant arc properties. They find that observationally based constraints suggest that there are not enough massive galaxies in a cluster to significantly alter arc statistics. In a complementary study, Menghetti et al. (2000) artificially added galaxies to N-body simulations. When the properties of short arcs are excluded, their analysis suggests that the contribution of cluster galaxies is negligible. 

A proper account of the cluster density profile in the presence of a centrally dominant galaxy requires a full treatment of galaxy formation at the center of the cluster. The studies mentioned artificially add intrinsically or extrinsically truncated parametric galactic halos to the cluster halo. Williams, Navarro \& Bartelmann (1999) (following Navarro, Frenk \& White (1997)) take the additional step of modifying the cluster halo adiabatically in response to the formed central galaxy. Profiles with large and small $\beta$ describe $CDM$ and $SIDM$ halos respectively. The response of the dark matter to the formation of a central galaxy cannot be commonly approximated in these two cases. 

In CDM halos, the central galaxy may form adiabatically (see Blumenthal, Faber, Flores \& Primack 1984), resulting in additional contraction of the cluster dark matter halo. In this scenario, it is possible that the central galaxy may retain its own halo if it is not tidally striped during formation. Note however that N-body simulations of cluster halos are expected to properly describe the halo of the central galaxy, since the cusp in the cluster density profile is formed from the infall of smaller halos. 

To reconcile observations with CDM NFW halo models, Williams et al. (1999) find that an additional central cluster galaxy of mass $\sim 3\times10^{12}M_{\odot}$is required. With a stellar velocity dispersion of $\sigma_*=300\,km\,sec^{-1}$, this implies that the central galaxy must retain its halo during formation. To find an approximate contribution to the multiple imaging cross-section we have taken the most simple approach and superimposed the dark matter halo profile of Brainerd, Blandford \& Smail (1996), 
\begin{equation}
\rho(r) = \frac{\sigma_*^2r_o^2}{4\pi G r^2(r^2 + r_o^2)}, 
\end{equation}
onto the center of the Zhao profile. Here $\sigma_*$ is the central velocity dispersion and $r_o$ is the characteristic outer scale on the center of the Zhao profile. The profile has a total mass
\begin{equation}
M_{g}=\frac{\pi r_o\sigma_*^2}{2G}.
\end{equation}
We have calculated cross-sections to multiple imaging for Zhao profiles having $M_{200}\sim 10^{15}M_{\odot}$ in the presence of an additional central galaxy of mass $M_{g}=3\times10^{12}M_{\odot}$ with a central stellar velocity dispersion of $\sigma_*=300\,km\,sec^{-1}$. The resulting cross-section calculations are applicable to Zhao profiles with large $\beta$ and $C_{NFW}$ appropriate for CDM halos. The left hand panels of Tabs.~\ref{CDtab} and \ref{CDtabrel} show values of the cross-section to multiple imaging $\frac{\tau_{d+b}(z_s)}{\tau_{SIS}(z_s)}$, and the fractional increase in cross-section $\frac{\tau_{d+b}(z_s)}{\tau(z_s)}$ over a small grid of $\beta$ and $C_{NFW}$.

In the self interacting scenario, the response of the halo to the formation of the central galaxy should be larger. The contraction heats up the dark matter, after which interactions transfer the heat outwards, resulting in cooling and additional halo contraction. An isothermal approximation is therefore appropriate for the response of the SIDM halo to the formation of a central galaxy. Consider an initial dark matter density profile resulting from an isothermal distribution function:
\begin{equation}
\rho_d(r)\propto\exp{\frac{\Phi_d(r)}{kT}}
\end{equation}
where $k$ is Boltzmann's constant and $T$ the temperature. The ratio of the dark matter density before and after the addition of baryons is
\begin{equation}
\frac{\rho_{d+b}}{\rho_{d}} = \exp{\frac{\Phi_d-\Phi_{d+b}}{kT}}\sim\exp{\frac{2\sigma_*}{\sigma_{cstr}}}.
\end{equation}
Thus the density increase of the cluster halo with $\sigma_{cstr}\sim1000\,km\,sec^{-1}$ and central galaxy velocity dispersion $\sigma_*\sim300\,km\,sec^{-1}$, the density increase is of order 20\%. This density increase is negligible with respect to the orders of magnitude which separate the densities of different profiles in the inner most regions (see Fig.~\ref{profsurf}).

If the cluster has a SIDM halo, the central galaxy cannot have retained its own halo during formation. The additional galactic mass should therefore consist only of stars, and is thus more confined. We have calculated cross-sections to multiple imaging for Zhao profiles having $M_{200}\sim 10^{15}M_{\odot}$ in the presence of an additional central of mass $M_{g}=3\times10^{11}M_{\odot}$ (appropriate for the mass in stars of a typical central galaxy) with a central stellar velocity dispersion of $\sigma_*=300\,km\,sec^{-1}$. The calculation is applicable to all the Zhao profiles assuming that the central dark-matter distribution is described solely by the Zhao profile. The right hand panels of Tabs.~\ref{CDtab} and \ref{CDtabrel} show the resulting values of the cross-section to multiple imaging $\frac{\tau_{d+b}(z_s)}{\tau_{SIS}(z_s)}$, and the fractional increase in cross-section $\frac{\tau_{d+b}(z_s)}{\tau(z_s)}$ where a central galaxy is included over a small grid of $\beta$ and $C_{NFW}$. $\frac{\tau_{d+b}(z_s)}{\tau(z_s)}$ is of order 1 for all but the shallowest halo profiles.

The addition of the extra mass increases the cross-section of all profiles with respect to the SIS. We have not re-normalized the mass of the Zhao-profile-plus-galaxy as we are interested in the relative rather than the absolute cross-sections of different Zhao-profile-plus-galaxies. The addition of the central isothermal galaxy sets a lower limit on the cross-section, and reduces the dependence of multiple imaging on the cluster profile parameters, though the differences are still orders of magnitude. The values in Tab.~\ref{CDtab} may be compared directly with the magnification bias (Tabs.~\ref{abbright} and \ref{abfaint}). These show that the contribution of the central galaxy to the optical depth is not as important as that of magnification bias. Note that while magnification bias and the addition of a central galaxy both preferentially increase the cross-section of less concentrated clusters, the addition of a central galaxy reduces the magnification bias due to the increased cross-section. Thus, while a self-consistent calculation is yet to be done, we conclude that centrally dominant galaxies will not prevent lens statistics from being used as a probe of SIDM halos.

\section{CONCLUSIONS}

We have calculated the differential and total optical depths to multiple imaging, the average image splitting and the total magnification for a constant co-moving number density of unevolving generalized NFW (Zhao) profile cluster mass gravitational lenses. We find that the number of expected strongly lensed quasars is a very sensitive function of the profile parameters. Profiles whose central density is low either due to a shallow central cusp, or to a scale radius that is a reasonable fraction of the virial radius have cross-sections to multiple imaging that are reduced by a significant factor. Moreover, the separation of multiple images is reduced (by a factor of a few), although the total magnification is significantly enhanced. We find that the resulting magnification bias does not alter our conclusions.

The Zhao profile exhibits degeneracies between profile parameters with respect to lensing statistics. Similarly, profile fits have parameter degeneracies which are a function of the minimization quantity adopted. If the properties of a profile core are accurately reproduced by an approximately degenerate profile, then the parameter degeneracies are nearly equivalent to those obtained from strong lensing statistics. However, a profile that attempts to fit the entire halo will introduce serious uncertainty into the inferred lensing rates (up to an order of magnitude). This is particularly true in the region of parameter space where both $C_{NFW}$ and $\beta$ are small and will be an important consideration for detailed calculations of the lensing rate for SIDM halos based on parametric results from N-body simulations. 

The lensing rate is a powerful probe of SIDM in clusters of galaxies, particularly in the absence of a centrally dominant galaxy. We have obtained profile parameters for halos around clusters of galaxies by fitting Zhao profiles to the simulations of Yoshida, Springel, White \& Tormen (2000). We find that the optical depth to multiple imaging seriously constrains the SIDM self-interaction cross-section. In particular, an interaction cross-section of $10.0cm^2gm^{-1}$ is rarely capable of producing multiple images during its evolution. In the most highly concentrated phases, a halo composed of interacting dark matter with this cross-section has a rate of multiple imaging 1 to 2 orders of magnitude lower than the corresponding typical CDM halo. Preliminary calculations show that centrally dominant galaxies increase the multiple imaging cross-section, but that the increase is small with respect to the variation in cross-section among different profiles. The presence of centrally dominant galaxies should not therefore inhibit the use of the multiple imaging rate as a probe of SIDM. 

An important result from this study, with implications for cluster lensing studies is that the scatter in CDM/SIDM profile parameters obtained from N-body studies describe a very large range of optical depth. The lensing statistics will therefore be dominated by the more concentrated members of the population, rather than by the typical halo. As a result, an estimate of the distribution of halo profiles must be included in future studies of cluster lensing statistics.

\acknowledgements
The authors would like to thank Romeel Dav$\acute{\mathrm{e}}$, Bartosz Pindor and Daniel Mortlock for helpful and stimulating discussions. We would also like to thank the anonymous referee whose comments led to the improvement of this work. This research was supported by NSF grant AST98-02802 to ELT. JSBW acknowledges the support of an Australian Postgraduate award and a Melbourne University Overseas Research Experience Award.

\clearpage

\figcaption[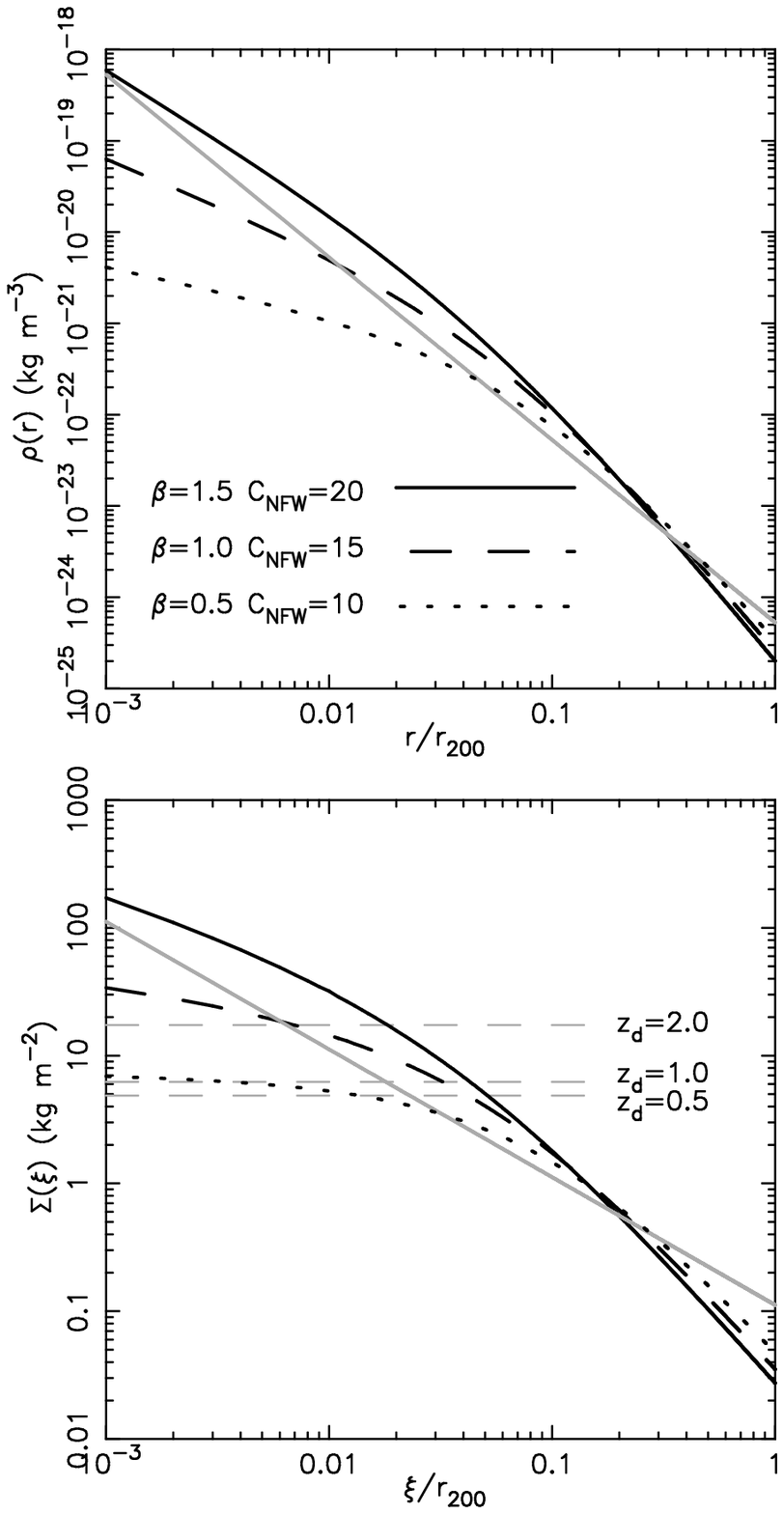]{Plots of $\rho(r)$ and $\Sigma(\xi)$ for Zhao profiles $I$ (solid lines), $II$ (dashed lines) and $III$ (dotted lines). $\rho_{SIS}(r)$ and $\Sigma_{SIS}(\xi)$ for a singular isothermal sphere are also shown (light lines). The dashed lines in the lower panel show the critical density for a source at $z_s=3$ and lenses at $z_d=0.5,1.0$ and $2.0$.\label{profsurf}}

\figcaption[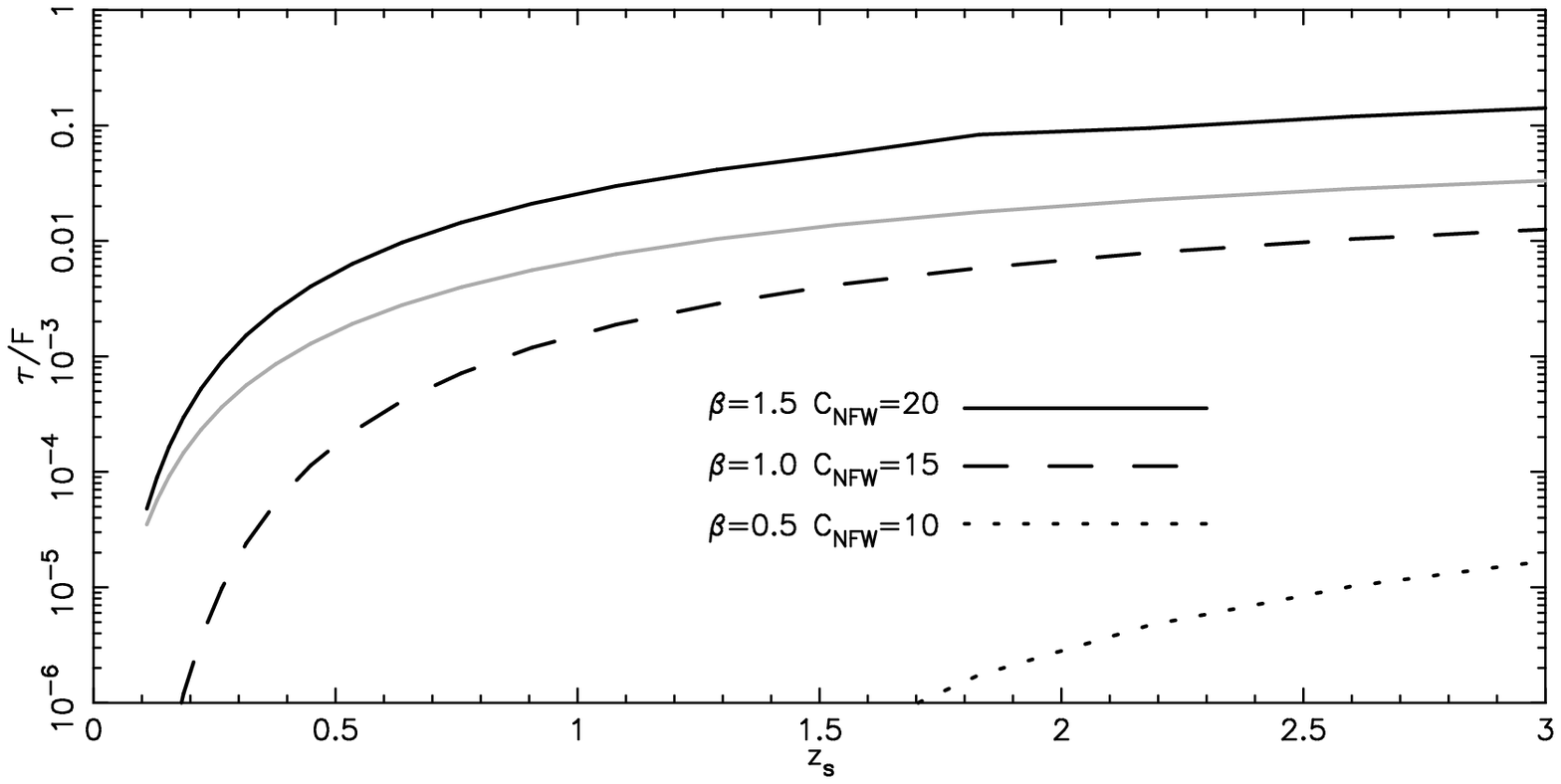]{Plots of $\tau(z_s)$ for Zhao profiles $I$ (solid line), $II$ (dashed line) and $III$ (dotted line). $\tau_{SIS}(z_s)$ for a singular isothermal sphere is also shown (light line). $\tau(z_s)$ has been normalized by $F_{SIS}$.\label{od}}

\figcaption[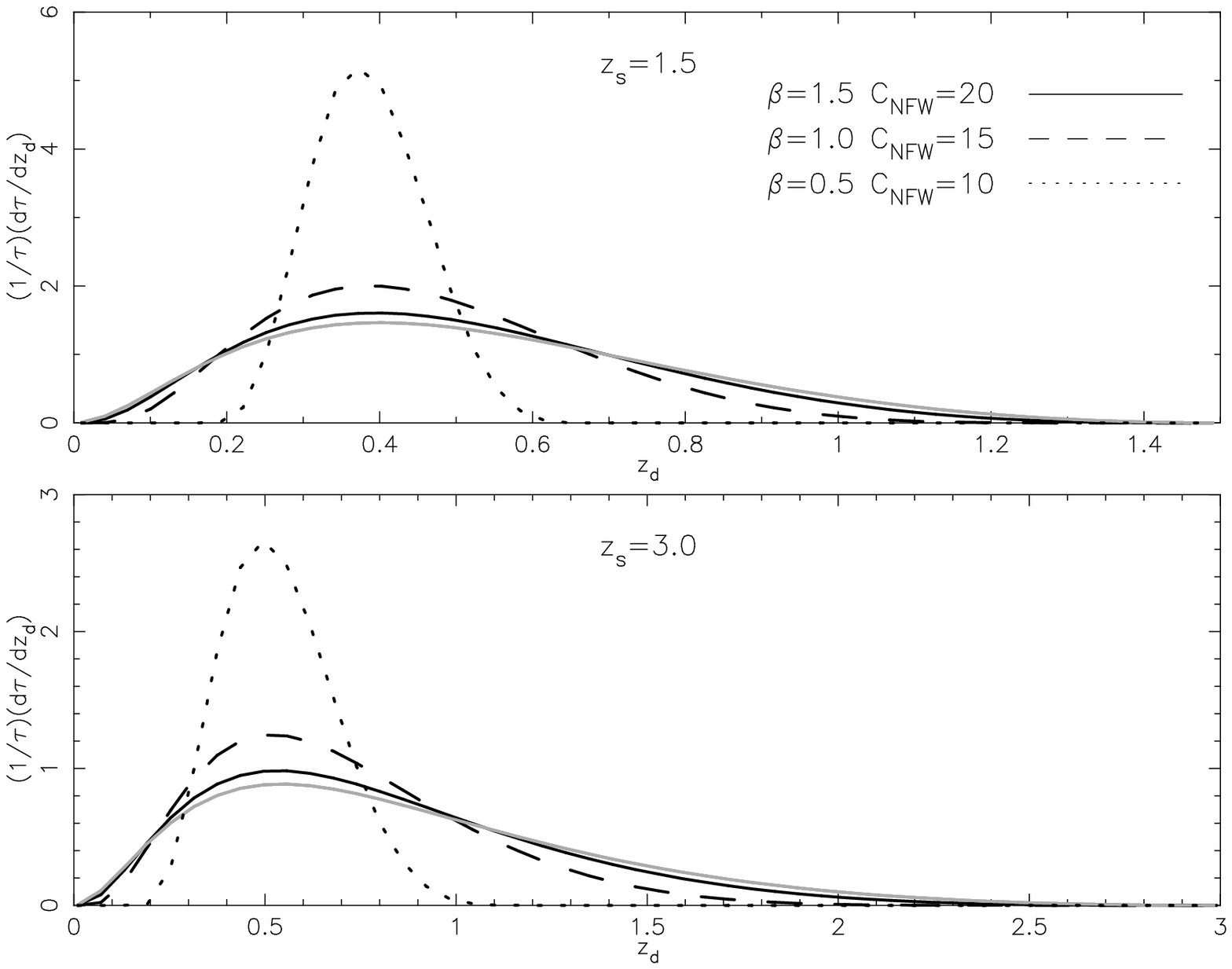]{Plots of $\frac{1}{\tau(z_s)}\frac{d\tau(z_s,z_d)}{d z_d}$ for profiles $I$ (solid lines), $II$ (dashed lines) and $III$ (dotted lines) with $z_s=1.5$ (upper panel) and $z_s=3.0$ (lower panel). The light lines show the corresponding distributions for a SIS.\label{difod}}

\figcaption[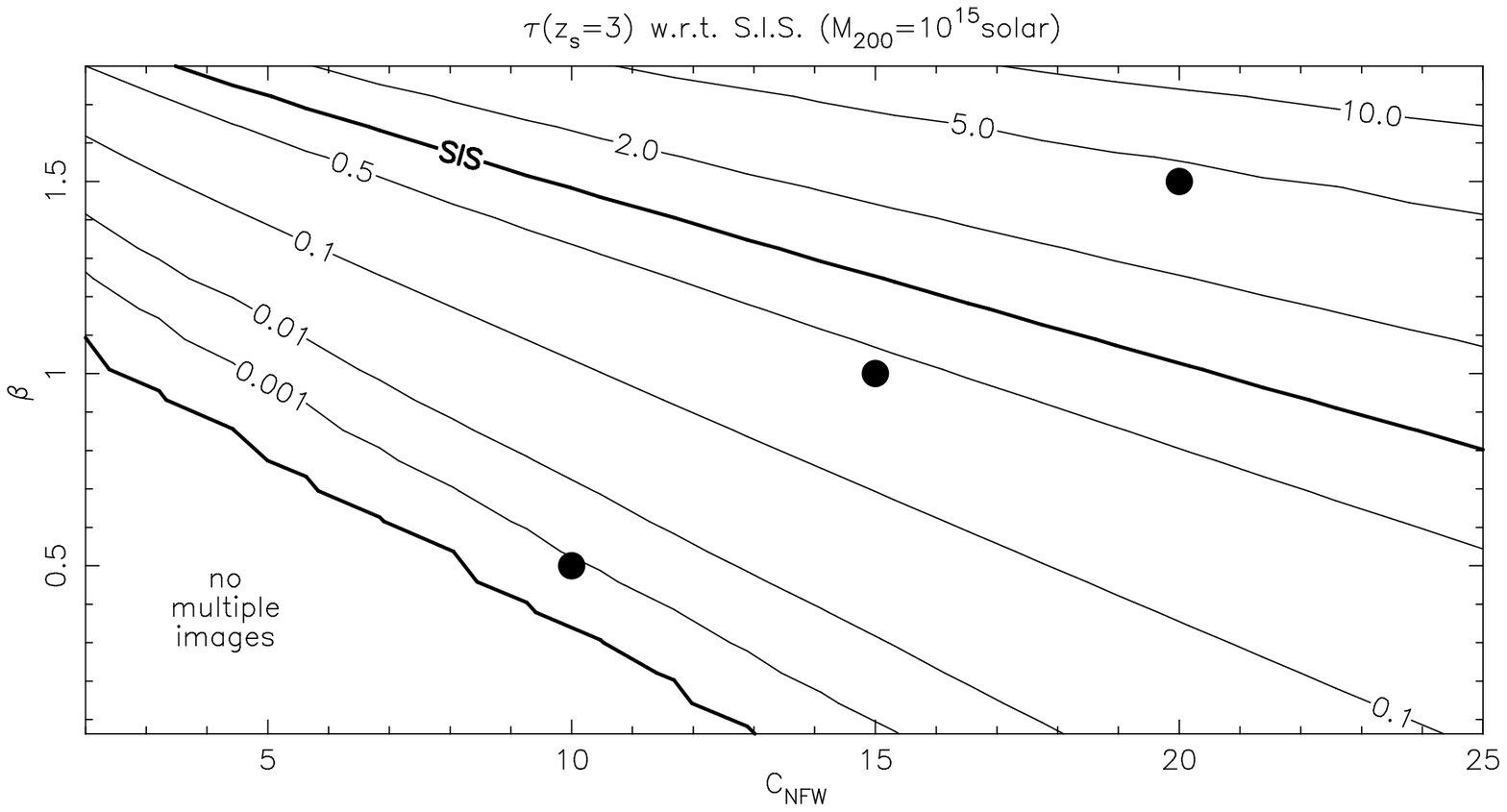]{Contour plot of $\frac{\tau(z_s)}{\tau_{SIS}(z_s)}$ with $z_s=3$ over a parameter space of $\beta$ and $C_{NFW}$. The upper thick line is a contour of unity and marks the parameters that produce an cross-section equal to that of a SIS. The lower thick line marks the boundary of the region of parameter space where no multiple imaging is produced. The dots show the locations of profiles $I$, $II$, and $III$ (left to right).\label{contod}}

\figcaption[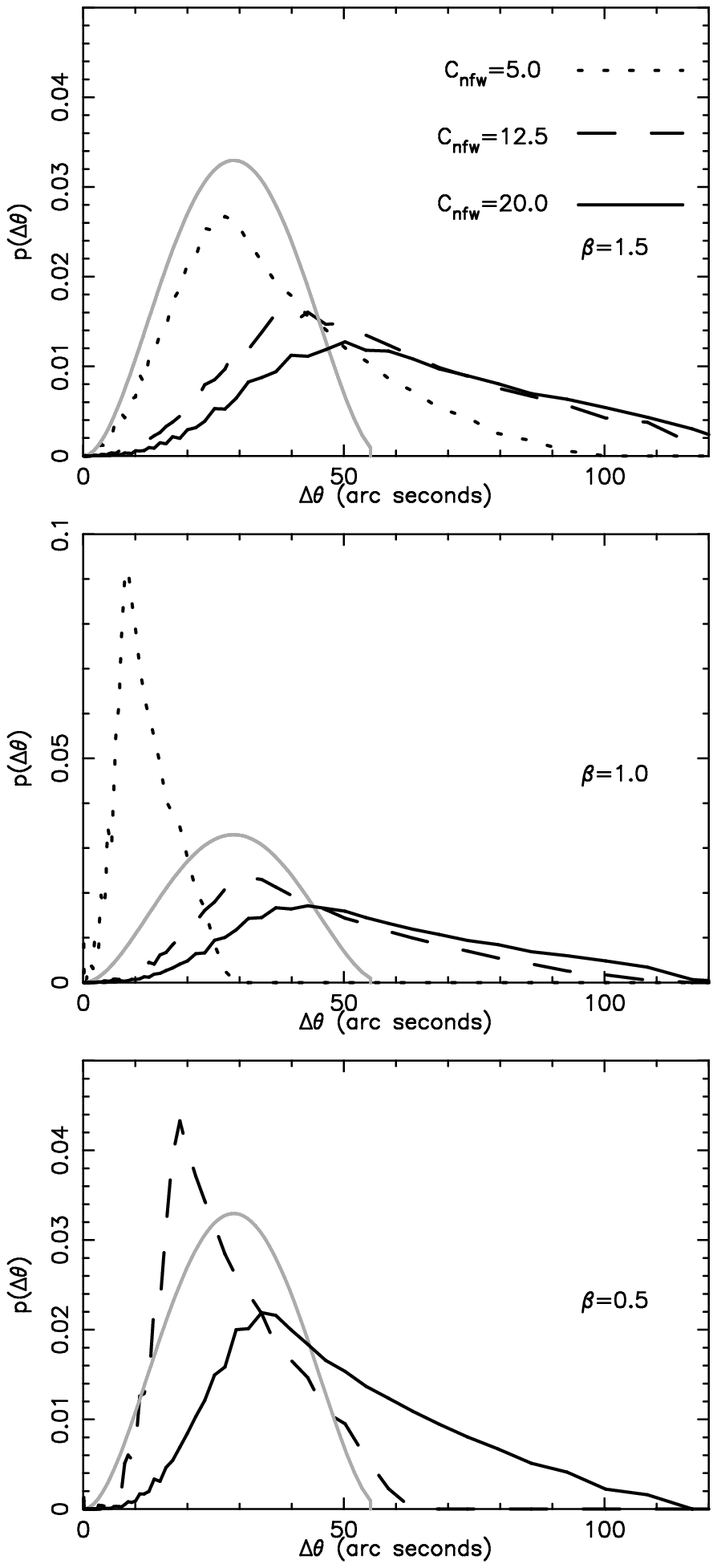]{Distributions of image separations over a grid of $C_{NFW}=$20.0 (solid lines), 12.5 (dashed lines), 5.0 (dotted lines) and $\beta=$1.5 (top), 1.0 (center), 0.5 (bottom) (note that $(C_{NFW},\beta)=(5.0,0.5)$ produces no multiple images and is not included). The light lines show the distribution for an isothermal sphere. \label{sep_dist}}

\figcaption[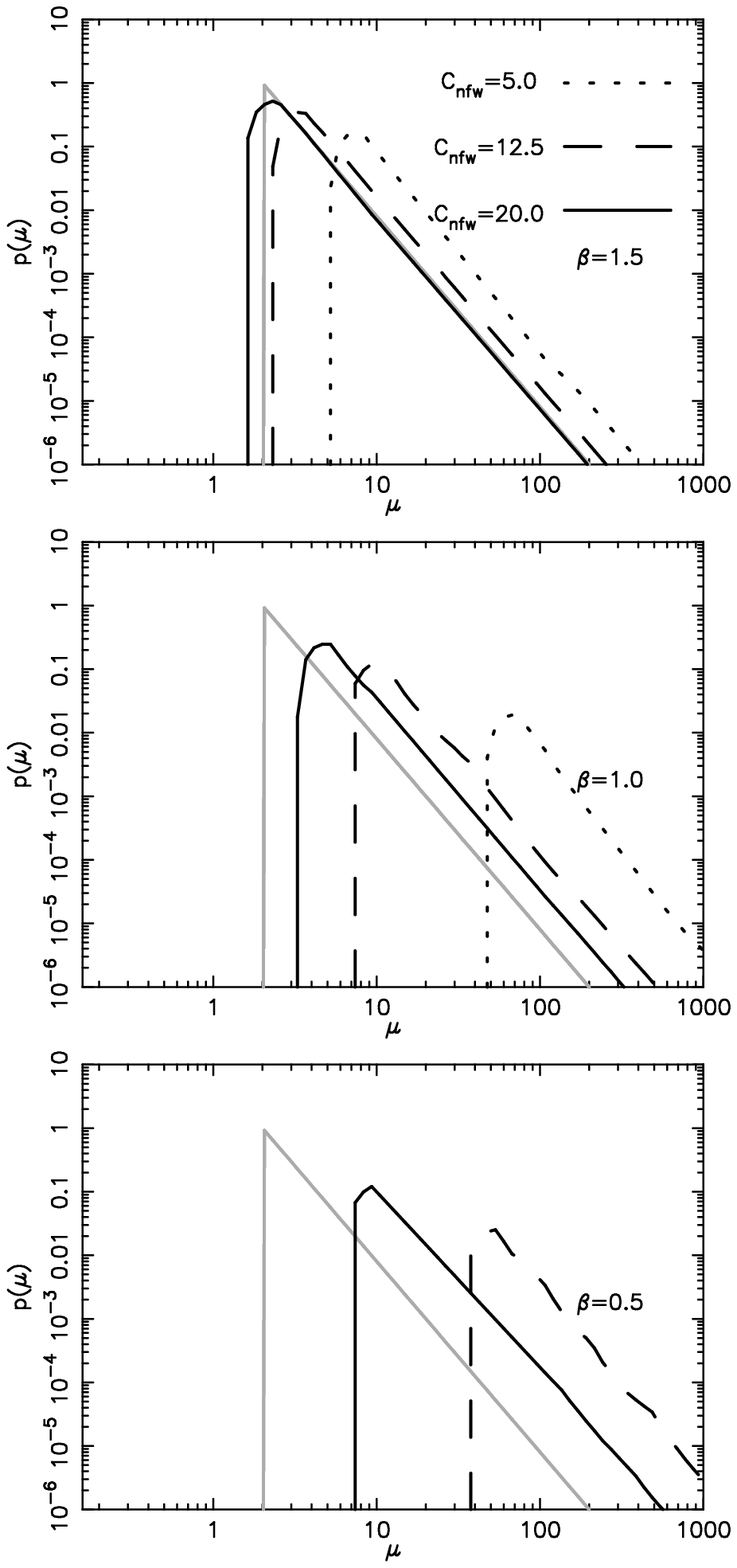]{Distributions of total image magnifications over a grid of $C_{NFW}=$20.0 (solid lines), 12.5 (dashed lines), 5.0 (dotted lines) and $\beta=$1.5 (top), 1.0 (center), 0.5 (bottom) (note that $(C_{NFW},\beta)=(5.0,0.5)$ produces no multiple images and is not included). The light lines show the magnification distribution for an isothermal sphere. \label{amp_dist}}

\figcaption[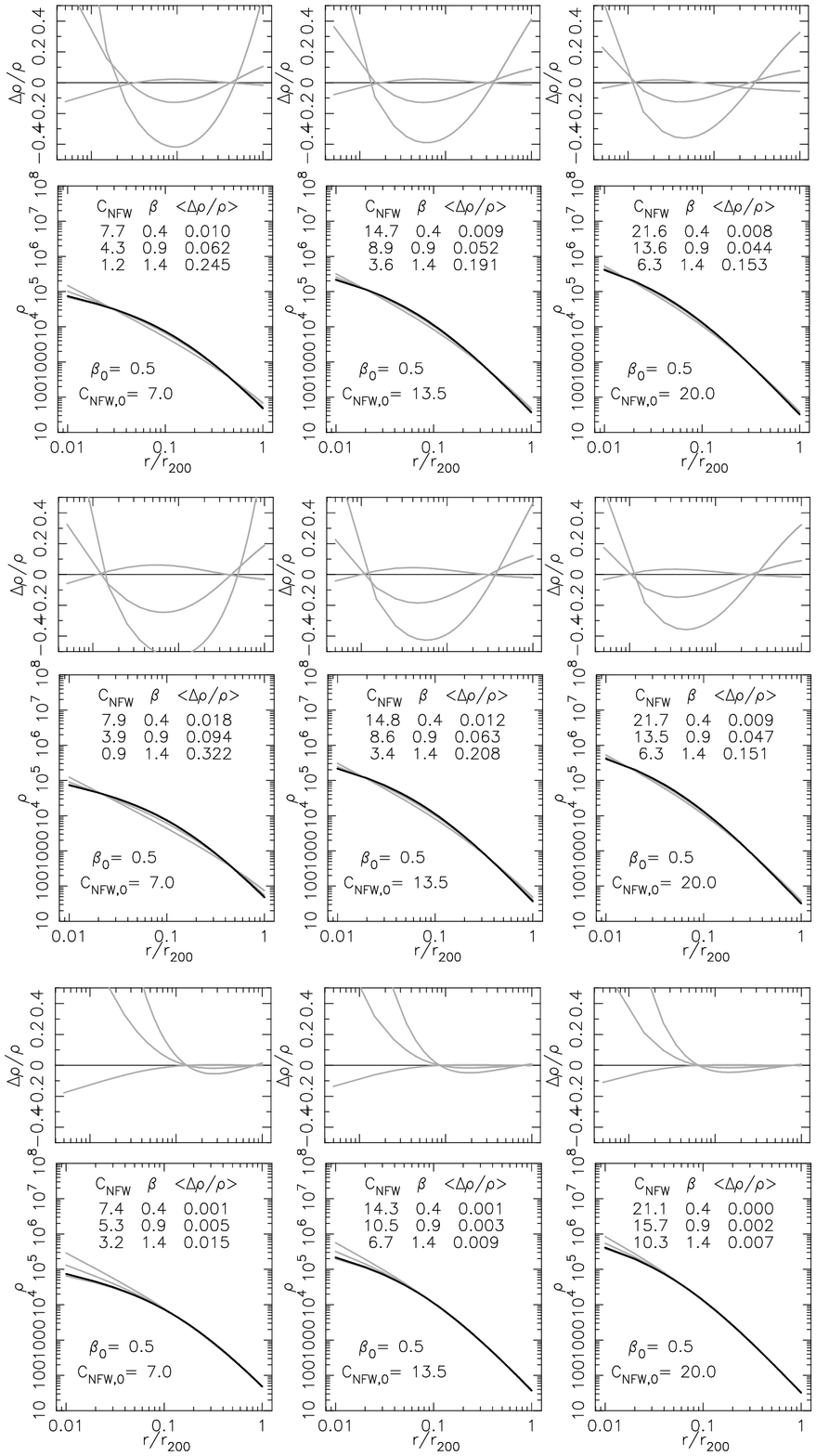]{A representation of the degeneracy between different Zhao profiles. Profiles with fixed $\beta$ were fit to a Zhao profile ($\beta_o=0.5$) over a range of concentration parameters $C_{NFW,o}$. The fitted $C_{NFW}$ and average fractional difference in density for a 3-d halo are shown. The fractional residuals are plotted above each set of fits. The minimization used the following functions. Top panels:  $\chi^2=\int_{\frac{r_{200}}{100}}^{r_{200}}\ r^2(\rho(r)-\rho_o(r))^2 dr$. Center panels: $max(|log(\rho(r))-\log(\rho_o(r))|)$. Bottom panels: $\chi_{rel}^2=\int_{\frac{r_{200}}{100}}^{r_{200}}\ r^2(\frac{\rho(r)-\rho_o(r)}{\rho_o(r)})^2 dr.$\label{fitprofa}}

\figcaption[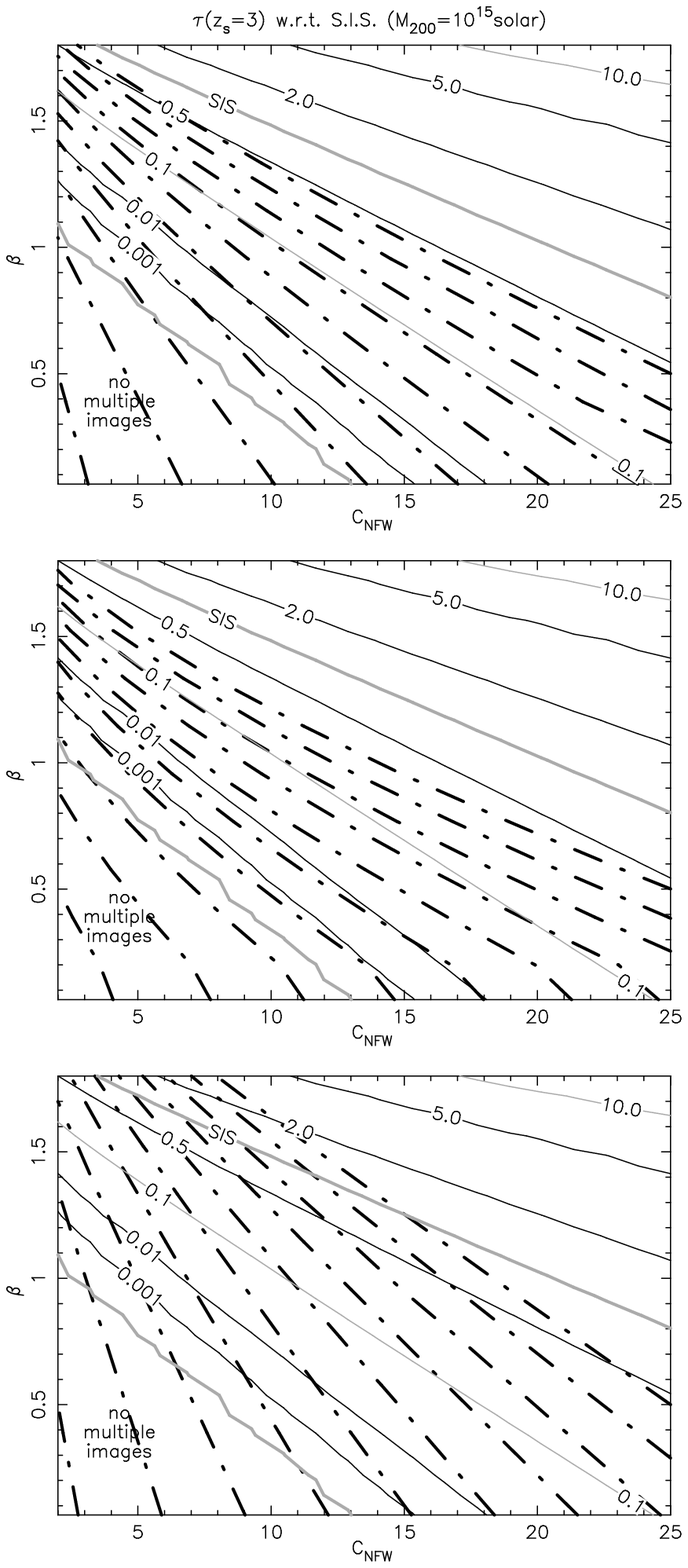]{A representation of the degeneracy between different Zhao profiles. The dot-dashed lines are curves along which the difference between halos with different parameters are minimized. The minimization used the following functions. Top figure:  $\chi^2=\int_{\frac{r_{200}}{100}}^{r_{200}}\ r^2(\rho(r)-\rho_o(r))^2 dr$. Center figure: $max(|log(\rho(r))-\log(\rho_o(r))|)$. Bottom figure: $\chi_{rel}^2=\int_{\frac{r_{200}}{100}}^{r_{200}}\ r^2(\frac{\rho(r)-\rho_o(r)}{\rho_o(r)})^2 dr$. The solid light lines in these figures represent the contours of $\frac{\tau(z_s)}{\tau_{SIS}(z_s)}$ with $z_s=3$. \label{fitprofb}}

\figcaption[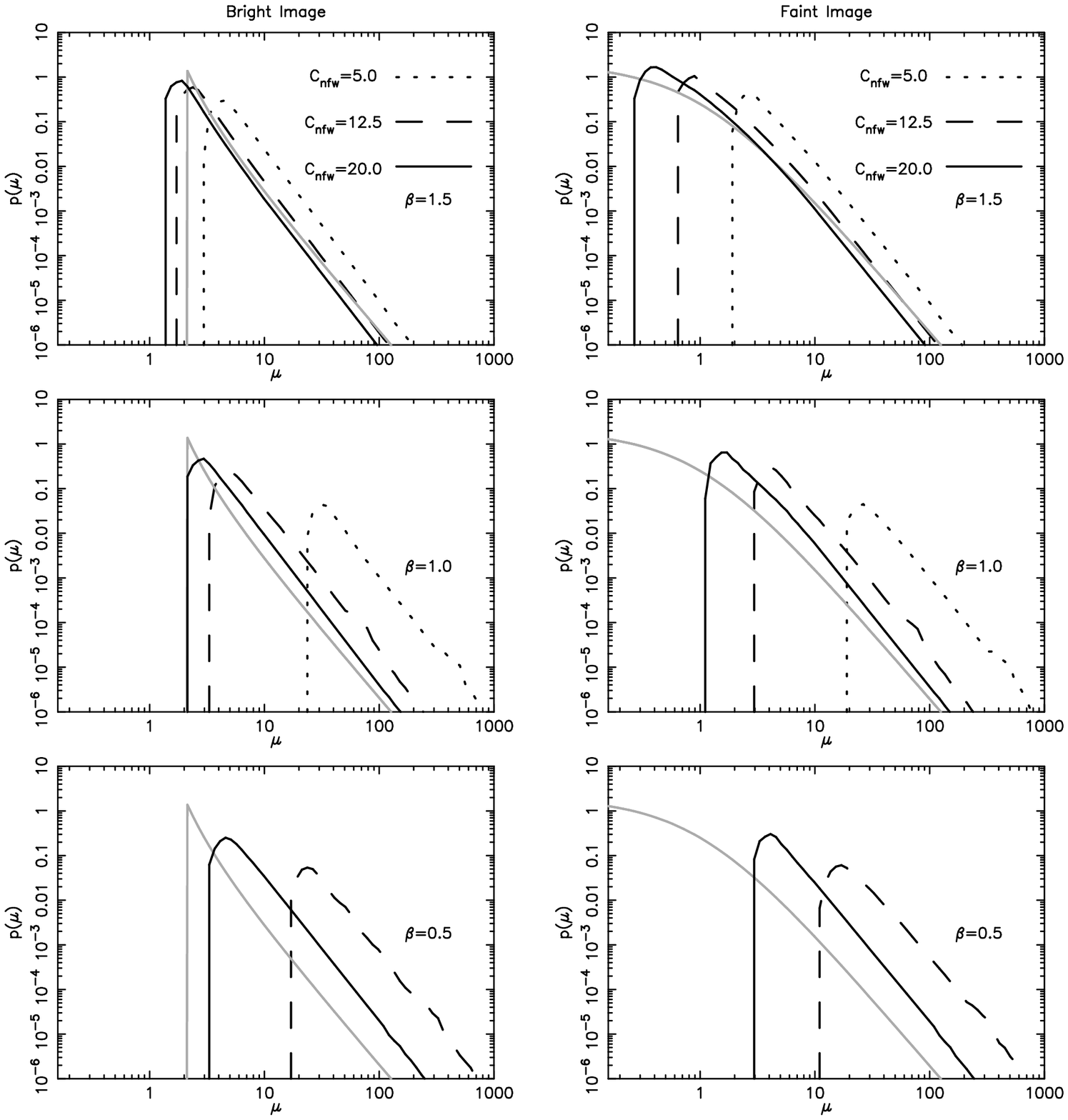]{Distributions of bright (left hand panels) and faint (right hand panels) image magnifications over a grid of $C_{NFW}=$20.0 (solid lines), 12.5 (dashed lines), 5.0 (dotted lines) and $\beta=$1.5 (top), 1.0 (center), 0.5 (bottom) (note that $(C_{NFW},\beta)=(5.0,0.5)$ produces no multiple images and is not included). The light lines show the corresponding bright and faint image magnification distributions for an isothermal sphere. \label{amp_dists_fig}}

\figcaption[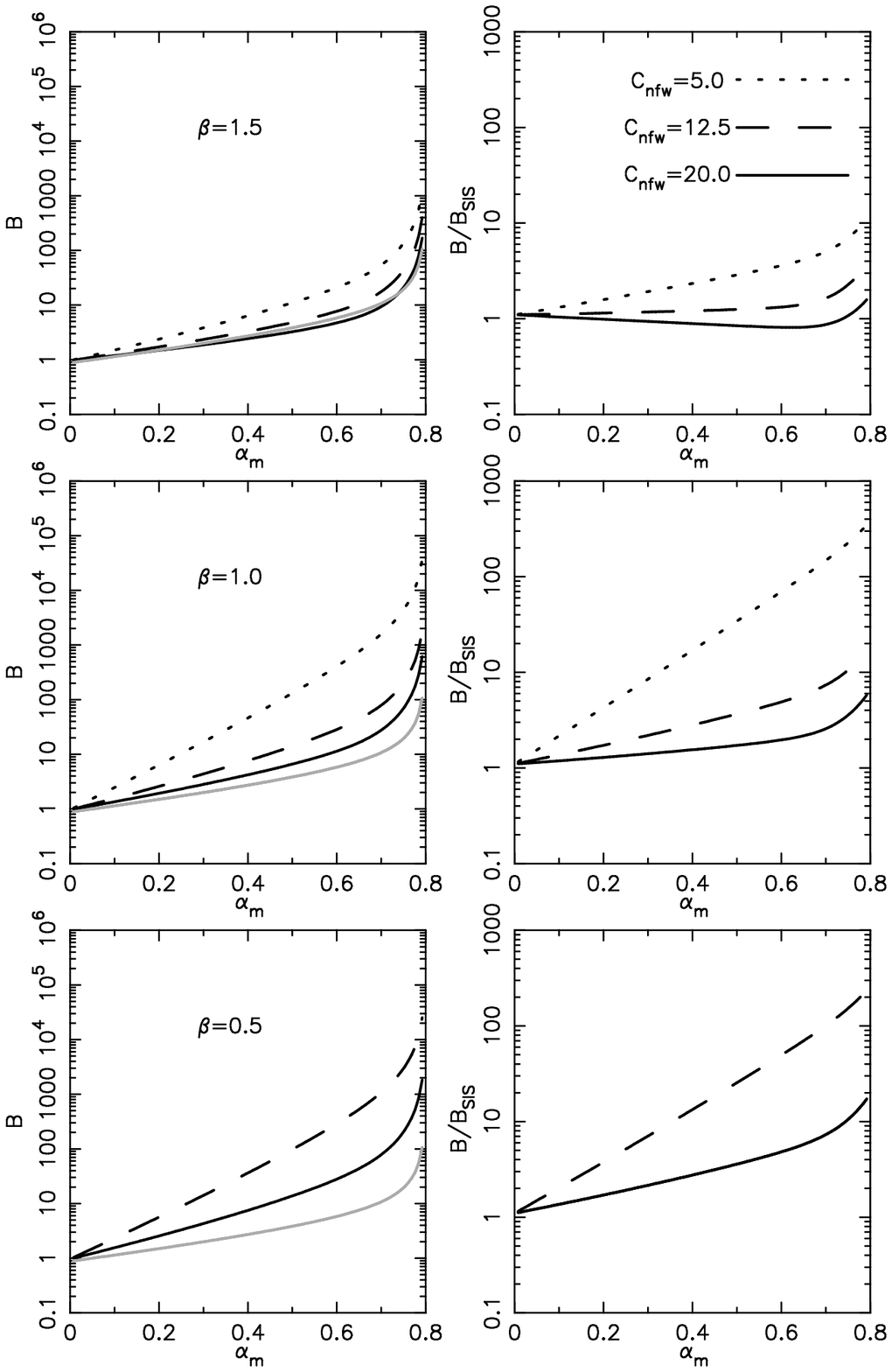]{The magnification bias for bright images as a function of $\alpha_m$ over a grid of $C_{NFW}=$20.0 (solid lines), 12.5 (dashed lines), 5.0 (dotted lines) and $\beta=$1.5 (top), 1.0 (center), 0.5 (bottom) (note that $(C_{NFW},\beta)=(5.0,0.5)$ produces no multiple images and is not included). Left: The absolute bias $B_b$. The light line shows the bias for an isothermal sphere $B_b^{SIS}$. Right: The magnification bias relative to the isothermal sphere $\frac{B_b}{B_b^{SIS}}$. \label{bright_amp_bias}}

\figcaption[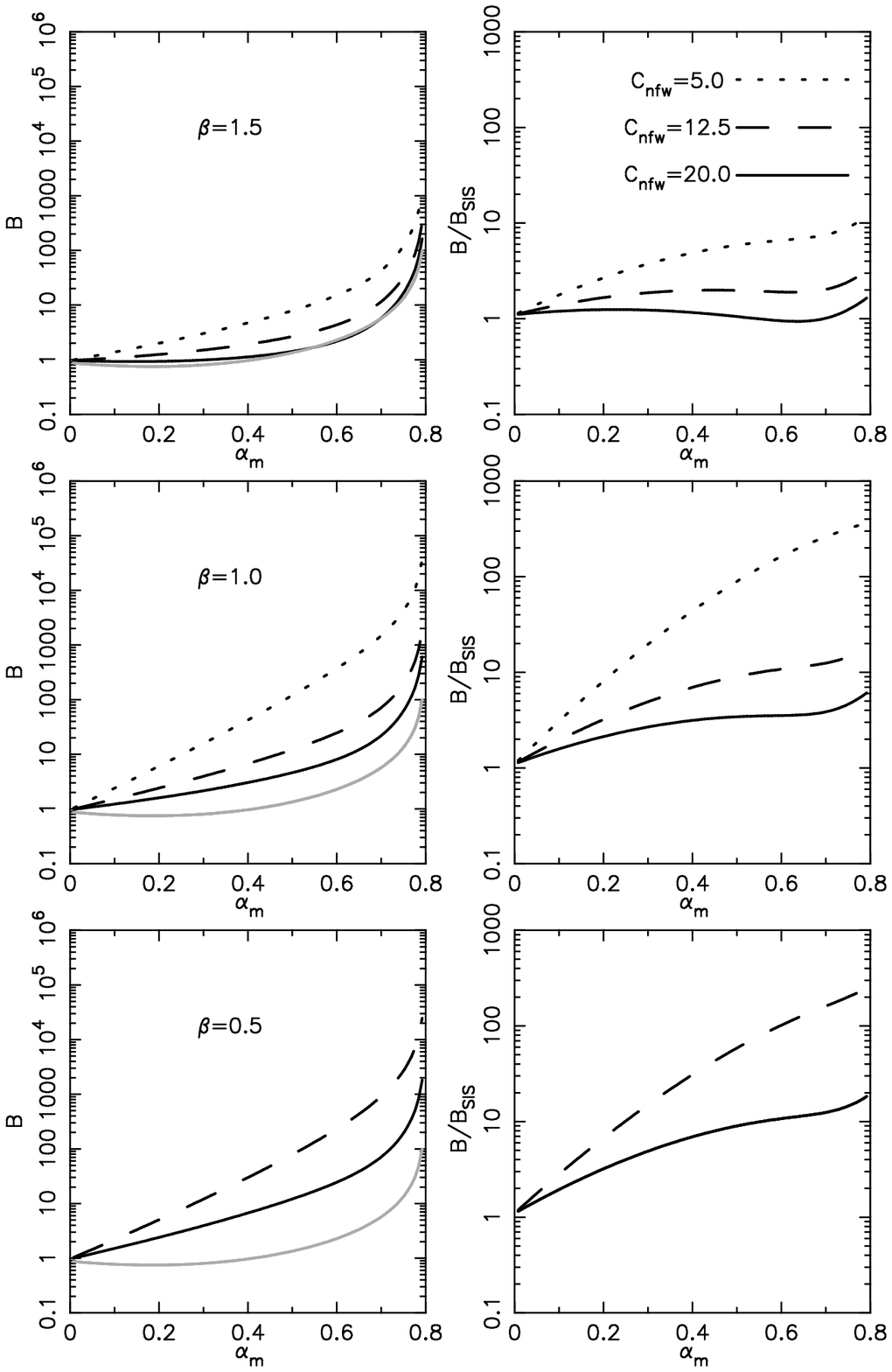]{The magnification bias for faint images as a function of $\alpha_m$ over a grid of $C_{NFW}=$20.0 (solid lines), 12.5 (dashed lines), 5.0 (dotted lines) and $\beta=$1.5 (top), 1.0 (center), 0.5 (bottom) (note that $(C_{NFW},\beta)=(5.0,0.5)$ produces no multiple images and is not included). Left: The absolute bias $B_f$. The light line shows the bias for an isothermal sphere $B_f^{SIS}$. Right: The magnification bias relative to the isothermal sphere $\frac{B_f}{B_f^{SIS}}$. \label{faint_amp_bias}}

\figcaption[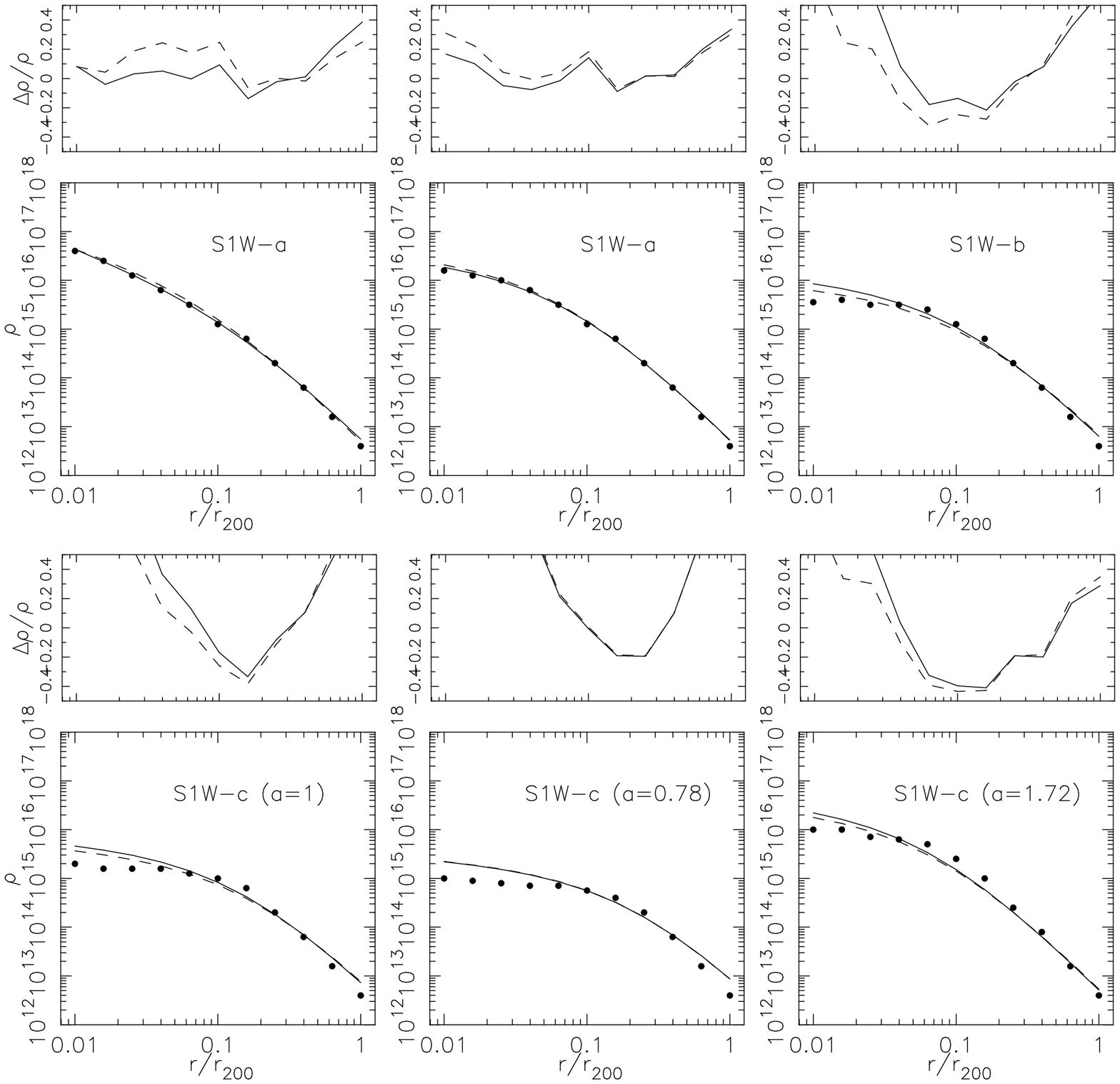]{Fits and residuals for the Yoshida et al. (2000) halos (dots). The solid and dashed lines correspond to minimization of $\chi^2$ and $\Delta\rho_{max}$. Each halo is labeled as per Yoshida et al. (2000).\label{yoshfit}}

\figcaption[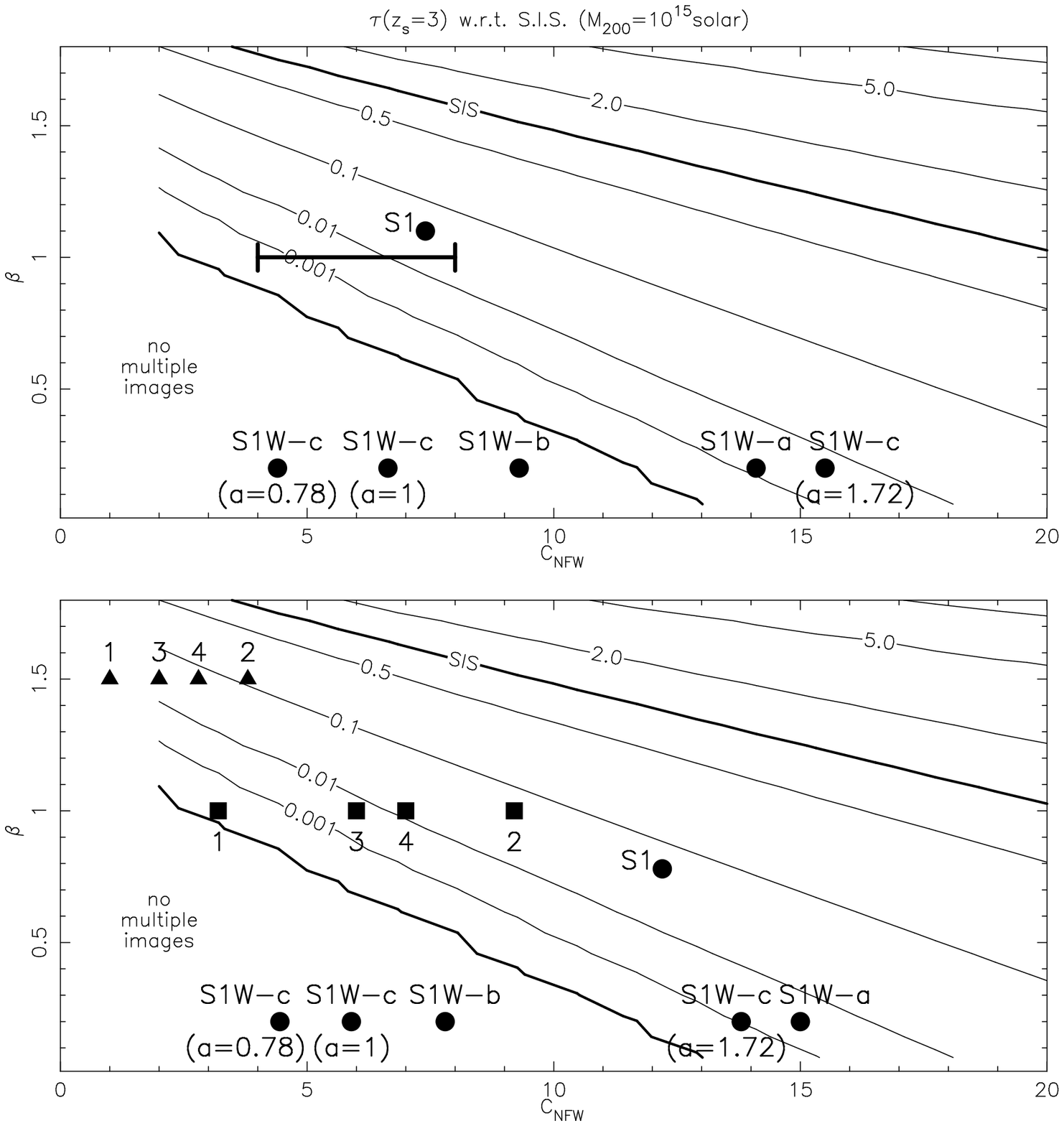]{Estimation of relative lensing rates for CDM and SIDM cluster halos. The dots represent the parameter combinations corresponding to fits for the Yoshida et al. (2000) halos (labeled, parameters in upper and lower plots used minimization of $\chi^2$ and $\Delta\rho_{max}$). The squares and triangles in the lower figure show the parameters (from Jing \& Suto 2000) resulting from fits of Zhao profiles with $\beta=1$ and $\beta=1.5$ respectively to CDM halos (labeled 1-4 as in Jing \& Suto 2000). The range bar in the upper figure shows the range (68\%) of concentrations found for the $\beta=1$ profile by Bullock et al. (2000) (extrapolated to $M_{200}\sim 10^{15}M_{\odot}$). These are superimposed on contours of $\frac{\tau(z_s)}{\tau_{SIS}(z_s)}$ with $z_s=3$.\label{yosh_points}}

\figcaption[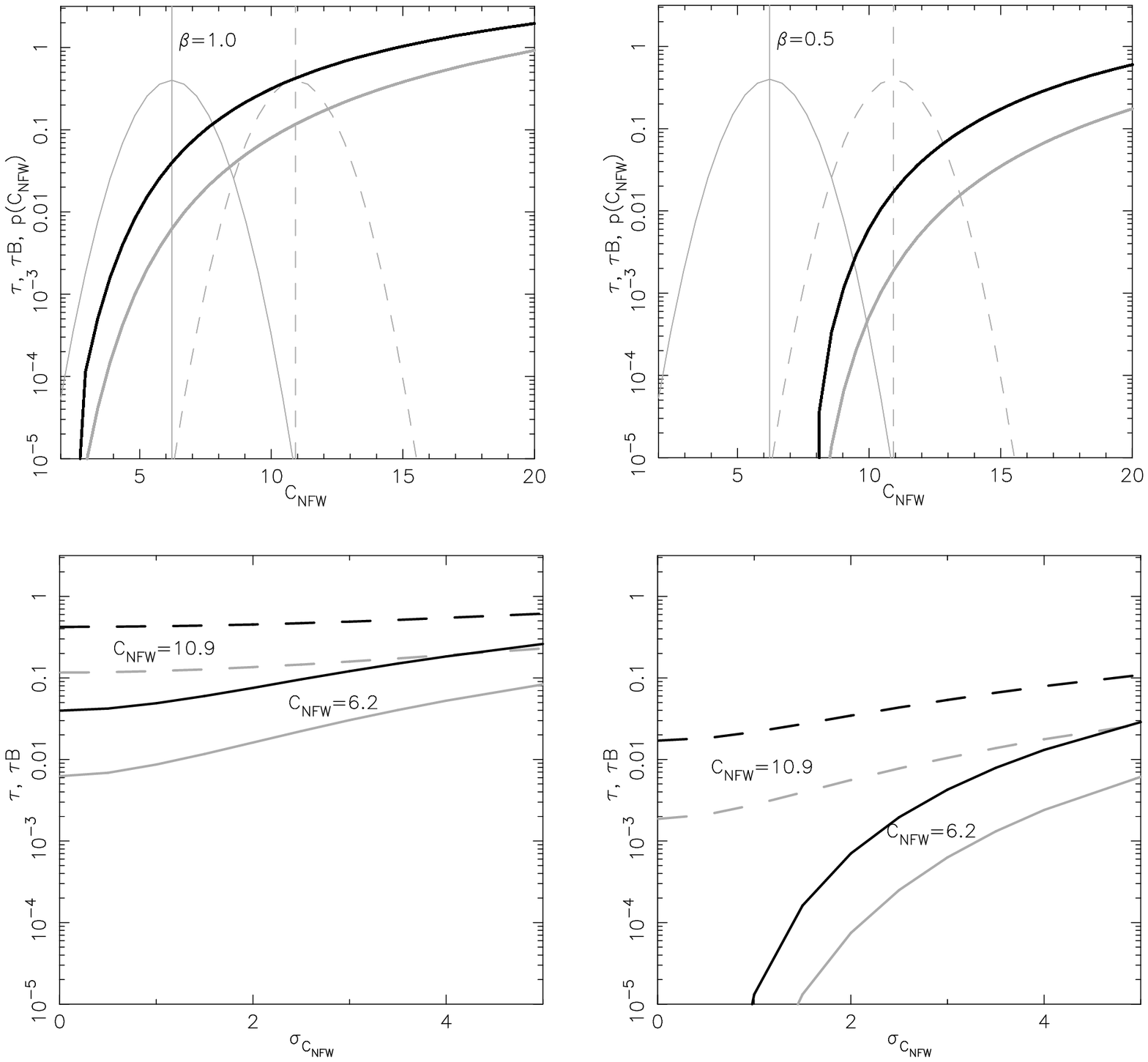]{Upper panels: $\frac{\tau}{\tau_{SIS}}$ (thick light lines) and $\frac{B_f}{B_f^{SIS}}\frac{\tau}{\tau_{SIS}}$ (thick dark lines) as a function of $C_{NFW}$ for $\beta=1.0$ (left hand panel) and $\beta=0.5$ (right hand panel). The thin solid and thin dashed curves show example $C_{NFW}$ distributions having $\langle C_{NFW}\rangle=6.1$ and $10.9$. In each case one delta function distribution and one Gaussian distribution (half-width of $\sigma_{C_{NFW}}=3$) are plotted. Lower panels: Convolutions of the cross-section $\frac{\tau}{tau_{SIS}}$ (light lines) and optical depth $\frac{B_f}{B_f^{SIS}}\frac{\tau}{\tau_{SIS}}$ (dark lines) with example Gaussian $C_{NFW}$ distributions as a function of $\sigma_{C_{NFW}}$. The solid and dashed lines correspond to convolutions with $C_{NFW}$ distributions having $\langle C_{NFW}\rangle=6.1$ and $10.9$ respectively. The Convolutions in the lower left and right panels assumed $\beta=1.0$ and $\beta=0.5$ respectively. We have assumed $\alpha_m=0.2$. \label{od_spread}}

\clearpage

\begin{figure}[htbp]
\centerline{\psfig{figure=fig0.eps,height=18cm}}
\end{figure}

\begin{figure}[htbp]
\centerline{\psfig{figure=fig1.eps,height=8cm}}
\end{figure}

\begin{figure}[htbp]
\centerline{\psfig{figure=fig2.eps,height=13cm}}
\end{figure}

\begin{figure}[htbp]
\centerline{\psfig{figure=fig3.eps,height=9cm}}
\end{figure}

\begin{figure}[htbp]
\centerline{\psfig{figure=fig4.eps,height=18cm}}
\end{figure}

\begin{figure}[htbp]
\centerline{\psfig{figure=fig5.eps,height=16cm}}
\end{figure}

\begin{figure}[htbp]
\centerline{\psfig{figure=fig6a.eps,height=195mm}}
\end{figure}

\begin{figure}[htbp]
\centerline{\psfig{figure=fig6b.eps,height=195mm}}
\end{figure}

\begin{figure}[htbp]
\centerline{\psfig{figure=fig7.eps,height=16cm}}
\end{figure}

\begin{figure}[htbp]
\centerline{\psfig{figure=fig8.eps,height=18cm}}
\end{figure}

\begin{figure}[htbp]
\centerline{\psfig{figure=fig9.eps,height=18cm}}
\end{figure}

\begin{figure}[htbp]
\centerline{\psfig{figure=fig10.eps,height=16cm}}
\end{figure}

\begin{figure}[htbp]
\centerline{\psfig{figure=fig11.eps,height=18cm}}
\end{figure}

\begin{figure}[htbp]
\centerline{\psfig{figure=fig12.eps,height=16cm}}
\end{figure}

\clearpage

\begin{table}[htbp]
\caption{{\small Values for $\frac{\tau(z_s)}{\tau_{SIS}(z_s)}$, $\langle\Delta\theta(z_s)\rangle$ and $\langle \mu(z_s)\rangle$ over a small grid of $\beta$ and $C_{NFW}$.}}
\vspace{3mm}
\begin{center}
\begin{tabular}{cccccccccccccc}
\cline{4-14}
       &   && \multicolumn{3}{c}{$\tau$ $(\tau_{SIS})$}&&\multicolumn{3}{c}{$\langle \Delta\theta\rangle$ (arc seconds)}&&\multicolumn{3}{c}{$\langle \mu\rangle$}\\\cline{4-6}\cline{8-10}\cline{12-14}
       &   && \multicolumn{3}{c}{$C_{NFW}$} &&\multicolumn{3}{c}{$C_{NFW}$} &&\multicolumn{3}{c}{$C_{NFW}$} \\

       &   && 5.0 &12.5 &20.0 && 5.0 &12.5 &20.0 && 5.0 &12.5 &20.0   \\\cline{1-2}\cline{4-6}\cline{8-10}\cline{12-14}
       &0.5&&0.000&0.008&0.174&&NA   &26.52&46.64&&NA   &75.41&18.46  \\
$\beta$&1.0&&0.001&0.199&0.918&&11.56&43.06&57.13&&100.1&16.11&8.105  \\ 
       &1.5&&0.226&1.715&4.263&&34.91&57.79&68.71&&12.33&5.732&3.861  \\\cline{1-2}\cline{4-6}\cline{8-10}\cline{12-14}

\end{tabular}
\end{center}
\label{sepmagtab}
\end{table}

\begin{table}[htbp]
\caption{{\small Values for the bright image magnification bias $B_b$ over a small grid of $\beta$ and $C_{NFW}$.}}
\vspace{3mm}
\begin{center}
\begin{tabular}{cccccccccccccc}
\cline{4-14}
       &   && \multicolumn{3}{c}{$\alpha_m=\frac{1}{5}$}&&\multicolumn{3}{c}{$\alpha_m=\frac{2}{5}$}&&\multicolumn{3}{c}{$\alpha_m=\frac{3}{5}$}\\\cline{4-6}\cline{8-10}\cline{12-14}
       &   && \multicolumn{3}{c}{$C_{NFW}$} &&\multicolumn{3}{c}{$C_{NFW}$} &&\multicolumn{3}{c}{$C_{NFW}$} \\

       &   && 5.0 &12.5 &20.0 && 5.0 &12.5 &20.0 && 5.0 &12.5 &20.0   \\\cline{1-2}\cline{4-6}\cline{8-10}\cline{12-14}
       &0.5&&NA   &5.53 &2.50 &&NA   &35.57&7.35 &&NA   &282.1&27.45  \\
$\beta$&1.0&&6.27 &2.53 &1.89 &&45.32&7.50 &4.13 &&398.1&27.91&11.18  \\ 
       &1.5&&2.32 &1.68 &1.45 &&6.24 &3.22 &2.38 &&20.50&7.54 &4.62   \\\cline{1-2}\cline{4-6}\cline{8-10}\cline{12-14}

\end{tabular}
\end{center}
\label{abbright}

\caption{{\small Values for the bright image magnification bias relative to the isothermal sphere $\frac{B_b}{B^{SIS}_b}$ over a small grid of $\beta$ and $C_{NFW}$.}}
\vspace{3mm}
\begin{center}
\begin{tabular}{cccccccccccccc}
\cline{4-14}
       &   && \multicolumn{3}{c}{$\alpha_m=\frac{1}{5}$}&&\multicolumn{3}{c}{$\alpha_m=\frac{2}{5}$}&&\multicolumn{3}{c}{$\alpha_m=\frac{3}{5}$}\\\cline{4-6}\cline{8-10}\cline{12-14}
       &   && \multicolumn{3}{c}{$C_{NFW}$} &&\multicolumn{3}{c}{$C_{NFW}$} &&\multicolumn{3}{c}{$C_{NFW}$} \\

       &   && 5.0 &12.5 &20.0 && 5.0 &12.5 &20.0 && 5.0 &12.5 &20.0   \\\cline{1-2}\cline{4-6}\cline{8-10}\cline{12-14}
       &0.5&&NA   &3.76 &1.70 &&NA   &13.34&2.76 &&NA   &49.51&4.82   \\
$\beta$&1.0&&4.27 &1.73 &1.29 &&16.99&2.81 &1.55 &&69.88&4.90 &1.96   \\ 
       &1.5&&1.57 &1.15 &0.99 &&2.34 &1.20 &0.89 &&3.60 &1.32 &0.81   \\\cline{1-2}\cline{4-6}\cline{8-10}\cline{12-14}

\end{tabular}
\end{center}
\label{abbrightSIS}
\end{table}

\begin{table}[htbp]
\caption{{\small Values for the faint image magnification bias $B_f$ over a small grid of $\beta$ and $C_{NFW}$.}}
\vspace{3mm}
\begin{center}
\begin{tabular}{cccccccccccccc}
\cline{4-14}
       &   && \multicolumn{3}{c}{$\alpha_m=\frac{1}{5}$}&&\multicolumn{3}{c}{$\alpha_m=\frac{2}{5}$}&&\multicolumn{3}{c}{$\alpha_m=\frac{3}{5}$}\\\cline{4-6}\cline{8-10}\cline{12-14}
       &   && \multicolumn{3}{c}{$C_{NFW}$} &&\multicolumn{3}{c}{$C_{NFW}$} &&\multicolumn{3}{c}{$C_{NFW}$} \\

       &   && 5.0 &12.5 &20.0 && 5.0 &12.5 &20.0 && 5.0 &12.5 &20.0   \\\cline{1-2}\cline{4-6}\cline{8-10}\cline{12-14}
       &0.5&&NA   &4.94 &2.35 &&NA   &29.23&6.55 &&NA   &227.5&24.00  \\
$\beta$&1.0&&5.94 &2.36 &1.56 &&41.98&6.59 &2.97 &&364.7&24.12&7.90   \\ 
       &1.5&&1.96 &2.22 &0.91 &&4.61 &1.87 &1.10 &&14.76&4.25 &2.13   \\\cline{1-2}\cline{4-6}\cline{8-10}\cline{12-14}

\end{tabular}
\end{center}
\label{abfaint}

\caption{{\small Values for the faint image magnification bias relative to the isothermal sphere $\frac{B_f}{B^{SIS}_f}$ over a small grid of $\beta$ and $C_{NFW}$.}}
\vspace{3mm}
\begin{center}
\begin{tabular}{cccccccccccccc}
\cline{4-14}
       &   && \multicolumn{3}{c}{$\alpha_m=\frac{1}{5}$}&&\multicolumn{3}{c}{$\alpha_m=\frac{2}{5}$}&&\multicolumn{3}{c}{$\alpha_m=\frac{3}{5}$}\\\cline{4-6}\cline{8-10}\cline{12-14}
       &   && \multicolumn{3}{c}{$C_{NFW}$} &&\multicolumn{3}{c}{$C_{NFW}$} &&\multicolumn{3}{c}{$C_{NFW}$} \\

       &   && 5.0 &12.5 &20.0 && 5.0 &12.5 &20.0 && 5.0 &12.5 &20.0   \\\cline{1-2}\cline{4-6}\cline{8-10}\cline{12-14}
       &0.5&&NA   &6.73 &3.20 &&NA   &30.90&6.93 &&NA   &101.7&10.74  \\
$\beta$&1.0&&8.10 &3.21 &2.13 &&43.84&6.96 &3.14 &&163.1&10.79&3.53   \\ 
       &1.5&&2.67 &1.66 &1.24 &&4.88 &1.98 &1.16 &&6.60 &1.90 &0.95   \\\cline{1-2}\cline{4-6}\cline{8-10}\cline{12-14}

\end{tabular}
\end{center}
\label{abfaintSIS}
\end{table}

\begin{table}[htbp]
\caption{{\small Values for the bright image magnification bias weighted cross-section relative to the isothermal sphere $\frac{B_{b}}{B^{SIS}_b}\frac{\tau}{\tau_{SIS}}$ over a small grid of $\beta$ and $C_{NFW}$.}}
\vspace{3mm}
\begin{center}
\begin{tabular}{cccccccccccccc}
\cline{4-14}
       &   && \multicolumn{3}{c}{$\alpha_m=\frac{1}{5}$}&&\multicolumn{3}{c}{$\alpha_m=\frac{2}{5}$}&&\multicolumn{3}{c}{$\alpha_m=\frac{3}{5}$}\\\cline{4-6}\cline{8-10}\cline{12-14}
       &   && \multicolumn{3}{c}{$C_{NFW}$} &&\multicolumn{3}{c}{$C_{NFW}$} &&\multicolumn{3}{c}{$C_{NFW}$} \\

       &   && 5.0 &12.5 &20.0 && 5.0 &12.5 &20.0 && 5.0 &12.5 &20.0   \\\cline{1-2}\cline{4-6}\cline{8-10}\cline{12-14}
       &0.5&&0.000&0.030&0.296&&0.000&0.107&0.480&&0.000&0.396&0.839  \\
$\beta$&1.0&&0.004&0.344&1.184&&0.017&0.559&1.423&&0.070&0.975&1.799  \\ 
       &1.5&&0.355&1.972&4.220&&0.529&2.058&3.794&&0.814&2.264&3.453  \\\cline{1-2}\cline{4-6}\cline{8-10}\cline{12-14}

\end{tabular}
\end{center}
\label{abbrighttao}

\caption{{\small Values for the faint image magnification bias weighted cross-section relative to the isothermal sphere $\frac{B_{f}}{B^{SIS}_f}\frac{\tau}{\tau_{SIS}}$ over a small grid of $\beta$ and $C_{NFW}$.}}
\vspace{3mm}
\begin{center}
\begin{tabular}{cccccccccccccc}
\cline{4-14}
       &   && \multicolumn{3}{c}{$\alpha_m=\frac{1}{5}$}&&\multicolumn{3}{c}{$\alpha_m=\frac{2}{5}$}&&\multicolumn{3}{c}{$\alpha_m=\frac{3}{5}$}\\\cline{4-6}\cline{8-10}\cline{12-14}
       &   && \multicolumn{3}{c}{$C_{NFW}$} &&\multicolumn{3}{c}{$C_{NFW}$} &&\multicolumn{3}{c}{$C_{NFW}$} \\

       &   && 5.0 &12.5 &20.0 && 5.0 &12.5 &20.0 && 5.0 &12.5 &20.0   \\\cline{1-2}\cline{4-6}\cline{8-10}\cline{12-14}
       &0.5&&0.000&0.054&0.557&&0.000&0.247&1.206&&0.000&0.814&1.869  \\
$\beta$&1.0&&0.008&0.639&1.955&&0.044&1.385&2.883&&0.163&2.147&3.240  \\ 
       &1.5&&0.603&2.847&5.286&&1.103&3.396&4.945&&1.492&3.259&4.050  \\\cline{1-2}\cline{4-6}\cline{8-10}\cline{12-14}

\end{tabular}
\end{center}
\label{abfainttao}
\end{table}

\begin{table}[htbp]
\caption{{\small Values for $\frac{\tau_{d+b}(z_s)}{\tau_{SIS}(z_s)}$, over a small grid of $\beta$ and $C_{NFW}$. Each cluster contains a central galaxy of mass $3\times10^{12}M_{\odot}$ (left) or $3\times10^{11}M_{\odot}$ (right) having a central velocity dispersion of $\sigma_{CD}=300\,km\,sec^{-1}$.}}
\vspace{3mm}
\begin{center}
\begin{tabular}{cccccccccc}
\cline{4-10}
       &   && \multicolumn{3}{c}{$M_g=3\times10^{12}M_{\odot}$}&&\multicolumn{3}{c}{$M_g=3\times10^{11}M_{\odot}$}\\\cline{4-6}\cline{8-10}
       &   && \multicolumn{3}{c}{$C_{NFW}$} &&\multicolumn{3}{c}{$C_{NFW}$} \\

       &   && 5.0 &12.5 &20.0 && 5.0 &12.5 &20.0 \\\cline{1-2}\cline{4-6}\cline{8-10}
       &0.5&&0.008&0.026&0.233&&0.008&0.015&0.188\\
$\beta$&1.0&&0.014&0.266&1.054&&0.010&0.217&0.952\\ 
       &1.5&&0.306&1.916&4.581&&0.257&1.778&4.364\\\cline{1-2}\cline{4-6}\cline{8-10}

\end{tabular}
\end{center}
\label{CDtab}

\caption{{\small Values for $\frac{\tau_{d+b}(z_s)}{\tau(z_s)}$, over a small grid of $\beta$ and $C_{NFW}$. Each cluster contains a central galaxy of mass $3\times10^{12}M_{\odot}$ (left) or $3\times10^{11}M_{\odot}$ (right) having a central velocity dispersion of $\sigma_{CD}=300\,km\,sec^{-1}$.}}
\vspace{3mm}
\begin{center}
\begin{tabular}{cccccccccc}
\cline{4-10}
       &   && \multicolumn{3}{c}{$M_g=3\times10^{12}M_{\odot}$}&&\multicolumn{3}{c}{$M_g=3\times10^{11}M_{\odot}$}\\\cline{4-6}\cline{8-10}
       &   && \multicolumn{3}{c}{$C_{NFW}$} &&\multicolumn{3}{c}{$C_{NFW}$} \\

       &   && 5.0 &12.5 &20.0 && 5.0 &12.5 &20.0 \\\cline{1-2}\cline{4-6}\cline{8-10}
       &0.5&&   NA&3.250&1.339&&   NA&1.875&1.080\\
$\beta$&1.0&&14.00&1.337&1.148&&10.00&1.090&1.037\\ 
       &1.5&&1.354&1.117&1.075&&1.137&1.037&1.024\\\cline{1-2}\cline{4-6}\cline{8-10}

\end{tabular}
\end{center}
\label{CDtabrel}
\end{table}

\end{document}